\documentclass[aps,prd,twocolumn,preprintnumbers,superscriptaddress,
              showpacs,nofootinbib]{revtex4}
\newcommand{\PRE}[1]{}				

\usepackage{amsmath}
\usepackage{amsfonts}
\usepackage{amssymb}
\usepackage{epsfig}
\usepackage{hyperref}
\usepackage{graphicx}

\newcommand{\alm}[1]{a_{l_#1 m_#1}}
\newcommand{\almn}{a_{l m}}
\newcommand{\bi}{B_{l_1 l_2 l_3}}
\newcommand{\bn}{\hat{\bf n}}
\newcommand{\deld}{\delta^{\rm D}}
\newcommand{\Ylm}[1]{Y_{l_#1}^{m_#1}}
\newcommand{\Ylmn}{Y_{l}^{m}}

\newcommand{\wjm}{\left(
                         \begin{array}{ccc}
       l_1 & l_2  & l_3  \\
         m_1 & m_2  & m_3
                         \end{array}
                   \right)}

\newcommand{\wj}{\left(
                         \begin{array}{ccc}
       l_1 & l_2  & l_3  \\
         0 & 0  & 0
                         \end{array}
                   \right)}

\newcommand{\ApJL}{Astrophys. J. Lett.}
\newcommand{\ApJ}{Astrophys. J.}
\newcommand{\PRL}{Phys. Rev. Lett.}

\begin{document}

\title{\PRE{\vspace*{1.5in}}
A Measurement of Primordial Non-Gaussianity Using WMAP 5-Year Temperature Skewness Power Spectrum
\PRE{\vspace*{0.3in}}}

\author{Joseph Smidt\footnote{jsmidt@uci.edu}}
\affiliation{Center for Cosmology, Department of Physics and Astronomy,
University of California, Irvine, CA 92697, USA }

\author{Alexandre Amblard\footnote{amblard@uci.edu}} 
\affiliation{Center for Cosmology, Department of Physics and Astronomy,
University of California, Irvine, CA 92697, USA }

\author{Paolo Serra\footnote{pserra@uci.com} }
\affiliation{Center for Cosmology, Department of Physics and Astronomy,
University of California, Irvine, CA 92697, USA }

\author{Asantha Cooray\footnote{acooray@uci.edu}} 
\affiliation{Center for Cosmology, Department of Physics and Astronomy,
University of California, Irvine, CA 92697, USA }

\date{\today}

\begin{abstract}
\PRE{\vspace*{.3in}}

We constrain the primordial non-Gaussianity parameter of the local model
$f_{\rm NL}$ using the skewness power spectrum associated with the two-to-one
cumulant correlator of cosmic microwave background temperature anisotropies.
This bispectrum-related power spectrum was constructed after weighting the
temperature map with the appropriate window functions to form an estimator
that probes the multipolar dependence of the underlying bispectrum associated
with the primordial non-Gaussianity. We also estimate a separate skewness power
spectrum sensitive more strongly to unresolved point sources.   When compared
to previous attempts at measuring the primordial non-Gaussianity with WMAP
data, our estimators have the main advantage that we do not collapse
information to a single number. When model fitting the two-to-one skewness power spectrum
we make use of bispectra generated by the primordial non-Gaussianity, radio point sources,
and lensing-secondary correlation. We analyze Q, V and W-band WMAP 5-year data using the KQ75 mask out to $l_{\rm
max}=600$.  Using V and W-band data and marginalizing over model parameters
related to point sources and lensing-secondary bispectrum, our overall and preferred constraint on $f_{\rm NL}$
is $11.0\pm23.7$ at the 68\% confidence level ($-36.4 < f_{\rm NL} < 58.4$ at
95$\%$ confidence).  We find no evidence for a non-zero
value of $f_{\rm NL}$ even marginally at the 1$\sigma$ level.

\end{abstract}


\maketitle


\section{Introduction}
\label{sec:intro}

The inflationary paradigm has deservedly become a cornerstone of modern
cosmology~\cite{Guth:1980zm,Linde:1981mu,Albrecht:1982wi,Starobinsky2,Sato:1980yn,Kazanas:1980tx}.  Inflation solves the
flatness, horizon and the monopole problems of the standard Big-Bang cosmology.
Furthermore, inflation is the prevailing paradigm related to the origin of
density perturbations that gave rise to the large-scale structure we see today.
It posits that a nearly exponential expansion stretched space in the first
moments of the early universe and promoted microscopic quantum fluctuations to
perturbations on cosmological scales today
\cite{GuthPi,Starobinsky,Hawking,Bardeen,Mukhanov}.  Inflation makes detailed
predictions for key statistical features of these fluctuations.  These
predictions have now begun to be tested by a range of cosmological
observations, including cosmic microwave background (CMB) temperature
anisotropy and polarization.

Recent measurements of the CMB with a variety of
ground, sub-orbital, and space-based experiments have provided some of the most
stringent tests of inflation (e.g., \cite{Komatsu:2008hk,Baumann:2008aq}).
Specifically among the generic predictions of inflation, recent CMB
measurements with the temperature anisotropy power spectrum and polarization
have established (1) a nearly flat geometry, (2) a nearly scale-invariant
spectrum at large angular scales, (3) adiabatic fluctuations, and (4)
super-horizon flucuations through the temperature-polarization cross spectrum.
One major prediction of inflation yet to be verified is the stochastic
background of primordial gravitational waves \cite{Abbott,Starobinsky2,Grishchuk:1974ny}.  While
strong limits are expected from Planck \cite{Planck}, a detection of the
gravitational wave background is the main focus of a next-generation
space-based  CMB experiment \cite{Bock:2006yf,Bock:2009xw,Bock:2008ww,Baumann:2008aj}.  

Some other tests of inflation involve the probability
distribution function and isotropy of the density perturbations generated by
inflation.  In the standard slow-roll inflationary model the inflaton, the
hypothesized scalar field or particle responsible for inflation, fluctuates with a minimal
amount of self interactions.  In fact, such a small amount of self interactions
ensures that the fluctuations are nearly Gaussian, and that any non-Gaussianity
produced would be too small for
detection \cite{Maldacena:2002vr,Acquaviva:2002ud,Sasaki:1995aw,Lyth:2004gb,Lyth:2005fi}.  Non-Gaussianity therefore
would be a measure of either interactions of the inflaton \cite{Allen87,Falk93}
or any non-linearities \cite{Salopek:1990jq,Gangui94}, and a detection  of
non-Gaussianity would indicate a violation of slow-roll inflation.  

In this spirit, models of non slow-roll inflation or alternatives to inflation
have been proposed to generate large, measurable non-Gaussianities.  The
curvaton mechanism produces curvature perturbations associated with the
fluctuations of a light scalar field whose energy density
is zero \cite{Mollerach:1989hu}.  The inhomogeneous reheating scenario can
produce non-Gaussianity through modulated reheating during the reheating stage
\cite{Dvali:2003em}.  Using multiple inflaton fields that are allowed to
interact, those interactions can be used to source non-Gaussianity
\cite{Linde:1984ti}.  Lastly, warm inflation~\cite{Berera:1995ie}, ghost
inflation \cite{ArkaniHamed:2003uy} and string theory inspired
D-cceleration~\cite{Silverstein:2003hf}  and Dirac-Born-Infeld (DBI) inflation~\cite{Chen:2005fe}
models also give rise to a large non-Gaussianity (see review in
Ref.~\cite{Bartolo:2004if}).

To connect with observable measurements, the associated non-Gaussianity of the
CMB can be described in terms of the second-order correction to the curvature
perturbations in position space with
\begin{equation}
\Phi({\bf x}) = \phi_L({\bf x}) + f_{\rm NL} \left[\phi_L^2({\bf x}) - \langle \phi_L({\bf x})\rangle^2\right] \, ,
\label{phi}
\end{equation}
where the non-Gaussianity parameter $f_{\rm NL}$ describes the amplitude of the
second-order correction.  This form was first suggested by Salopek \& Bond
\cite{Salopek:1990jq,Komatsu:2001rj} to describe the non-Gaussianity in primordial
perturbations from inflation and has been the subject of experimental
constraints using a variety of CMB and large-scale structure data in recent
years. 

Instead of constraints on the non-Gaussianity parameter in the position space,
recent studies make use of the bispectrum involving a three-point correlation
function in Fourier or multipole space.  The configuration dependence of the
bispectrum $B(k_1,k_2,k_3)$ with lengths $(k_1,k_2,k_3)$ that form a triangle
in Fourier space can be used to separate various mechanisms for
non-Gaussianities, depending on the effectiveness of of the estimator used.  To summarize the status of the non-Gaussianity measurements,
an analysis with WMAP 3-year data first suggested  a hint of a non-Gaussianity
in the local model with $27 < f_{\rm NL} < 147$ (95\% CL), far above the value
of $f_{\rm NL} < 1$ expected in simple, single field, slow-roll inflation
models \cite{Yadav:2007yy}.  The WMAP team's preferred measurement of
non-Gaussianity parameter in 5-year V and W-band data is 
$-9 < f_{\rm NL} < 111$ (95\% CL) \cite{Komatsu:2008hk}.  The most
recent constraint on $f_{\rm NL}$ comes from studying the WMAP 5-year data
with an optimal estimator  leading to $-4 < f_{\rm NL} < 80$
(95\% CL) \cite{Smith:2009jr}.  At the 68\% confidence level, with a value of
$f_{\rm NL}=38\pm 21$, there is still some marginal evidence for a non-zero
value of the non-Gaussianity parameter. If such a result were to continue to
hold with Planck, which increases the precision of $f_{\rm NL}$ measurement by a factor of 3 to 4,
then our simple inflationary picture would need to be revised to include a more complex model.

In this paper, we will pursue a new measurement of the primordial
non-Gaussianity parameter with a new estimator that preserves some
angular dependence of the bispectrum. On the contrary, the estimators
employed by most CMB non-Gaussianity studies, including those by the WAMP team
\cite{Komatsu:2008hk}, involves a measurement that compresses all information
of the bispectrum to a single number called the cross-skewness computed with
two weighted maps.  Such a drastic compression limits the ability to study the
angular dependence of the non-Gaussian signal and to separate any confusing
foregrounds from the primordial non-Gaussianity.  In addition to Galactic
foregrounds, non-Gaussianity measurements could also be contaminated by
unresolved point sources, mainly radio and dusty galaxies, and Sunyaev-Zel'dovich (SZ) clusters, among others
\cite{Serra}.  Given the increase in size of CMB data, especially with Planck,
it is also necessary to develop accurate measurement techniques to extract
$f_{\rm NL}$ that are unbiased.

Our estimator for non-Gaussianity uses a weighted version of the squared temperature-temperature angular power spectrum
\cite{Cooray:2001ps,Munshi:2009ik}, which we refer to as the skewness power
spectrum. This power spectrum extracts information from the bispectrum 
as a function of the multipole of one triangle length in the
harmonic space, while summing all configurations given by the other two side
lengths. The difference in spatial dependence based on how the maps are weighted provides
ways to separate primordial non-Gaussianity from that of the foregrounds.
Here, we account for both point source and lensing bispectra
with latter resulting from the correlation of the lensing potential with
secondary anisotropies \cite{Goldberg,CooHu}.

To summarize our main results, after marginalizing over the normalizations of
point source and lensing-secondary bispectra, with the combination of V and
W-band maps we are able to constrain $f_{\rm NL} = 11.0 \pm
23.7$ at the 68\% confidence level or $-36.4 < f_{\rm NL} < 58.4$ at the $95\%$
confidence level. We find that $f_{\rm NL}$ is never incompatible with zero at $68\%$
confidence when $f_{\rm NL}$ is estimated in independent bins 
of width 200 between $2 < l < 600$. We find a significant contribution from  unresolved
point sources, but failed to detect the lensing-secondary cross-correlations using the two statistics we considered
here. 

In section \S \ref{sec:theory} we review the background theory and in \S \ref{analysis}
we review the estimator used and our simulation procedure to compute the uncertainties.  In \S \ref{sec:data} we discuss our methods for analyzing and simulating data.
In section \S \ref{sec:results} we discuss our results.  In section
\S \ref{sec:conclusion} we conclude with a summary of our results.

\section{Theory}
\label{sec:theory}

To begin, we define multipole moments of the temperature map through
\begin{equation}
a_{lm} = \int d\bn T(\bn) \Ylmn {}^*(\bn)  \, .
\label{eqn:alm}
\end{equation}
The angular power spectrum and bispectrum are defined in the usual way such that
\begin{eqnarray}
\langle \alm{1}^* \alm{2}\rangle &=& \deld_{l_1 l_2} \deld_{m_1 m_2}
        C_{l_1}\,, \\
\langle \alm{1} \alm{2} \alm{3} \rangle  &=& \wjm \bi\, .
\end{eqnarray}
Here the quantity in parentheses is the Wigner-3$j$ symbol.
The orthonormality relation for Wigner-3$j$ symbol implies
\begin{eqnarray}
\bi = \sum_{m_1 m_2 m_3}  \wjm
                \langle \alm{1} \alm{2} \alm{3} \rangle \,.
\label{eqn:bispectrum}
\end{eqnarray}
The angular bispectrum, $\bi$, contains all the information available
from the three-point correlation function.  For example, the skewness,
the pseudo-collapsed 
three-point function of Ref.~\cite{Hinetal95}  and the
equilateral configuration statistic of Ref.~\cite{Feretal98}  can all
be expressed as linear combinations of the bispectrum
terms (see Ref.~\cite{Gangui94} for explicit expressions and
Ref.~\cite{Cooetal00} for an expression relating skewness in terms of
the bispectrum).

\begin{figure}
   \vspace{-0.6cm}
    \begin{center}
      \includegraphics[scale=0.47]{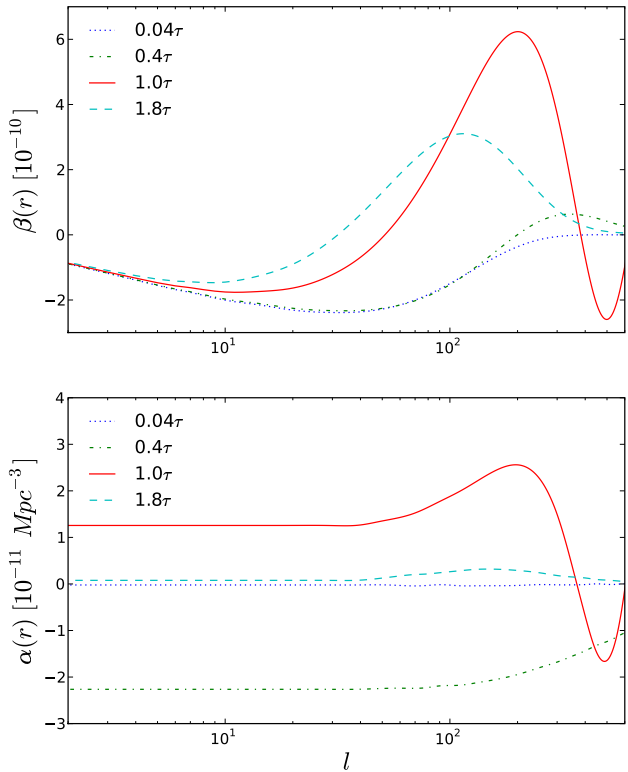} 
   \end{center}
   \vspace{-0.7cm}
   \caption[width=3in]{$\alpha_l(r)$ and $\beta_l(r)$ with respect to $l$ for $r$ values defined as followed: $r = c(\tau_{0}-a\tau)$ where $\tau_{0}$ is the present day conformal time and $c\tau = 235$ Mpc.  In these plots, a = 0.04, 0.4, 1.0 and 1.8.}
   \label{fig:alphabeta}
\end{figure}

\subsection{Primordial Non-Gaussianity}

Here we focus on the local form of the primordial non-Gaussianity. Using the second order correction to the curvature perturbations $\Phi$ in equation~(\ref{phi})
and following the derivation in Ref.~\cite{Komatsu:2001rj}, we write the
angular bispectrum of temperature anisotropies as
\begin{eqnarray}
 B^{\rm NG}_{l_1l_2l_3} &=& 2 I_{l_1l_2l_3}\int_0^\infty r^2 dr
  \left[\alpha_{l_1}(r)\beta_{l_2}(r)\beta_{l_3}(r) +
   (\mbox{Perm.})\right],\nonumber \\
\end{eqnarray}
where
\begin{equation}
 I_{l_1l_2l_3}\equiv \sqrt{\frac{(2l_1+1)(2l_2+1)(2l_3+1)}{4\pi}}
\left(\begin{array}{ccc}l_1&l_1&l_3\\0&0&0\end{array}\right) \, ,
\end{equation}
and $r$ is the comoving radial coordinate.

The two functions in $B^{\rm NG}_{l_1l_2l_3}$ are given by
\begin{eqnarray}
\label{eq:alpha}
 \alpha_l(r)&\equiv&\frac{2}{\pi}\int k^2dk g_{Tl}(k)j_l(kr),\\
 \label{eq:beta}
 \beta_l(r)&\equiv&\frac{2}{\pi}\int k^2dk P_\Phi(k)g_{Tl}(k)j_l(kr) \, .
\end{eqnarray}

Here, $P_\Phi(k)\propto k^{n_s-4}$ is the primordial power spectrum of
Bardeen's curvature perturbations, and $g_{Tl}(k)$ is the radiation transfer
function that gives the angular power spectrum as $C_l=(2/\pi)\int k^2 dk
P_\Phi(k)g_{Tl}^2(k)$.  In Fig.~1, we show four example cases of $\alpha(r)$
and $\beta(r)$.  We generate them using a modified version of the CMBFAST code
\cite{SelZal96} and for our fiducial cosmological parameter values, consistent with
WMAP 5-year best-fit model, as summarized in Table~I.

\begin{table}[htbp]
  \centering
  \begin{tabular}{@{} |c|c| @{}}
    \hline
    Parameter & Value \\ 
    \hline
    $H_0$ & $71.9 \ {\rm km/s/Mpc}$ \\ 
    $\Omega_b h^2$ & 0.02273 \\ 
    $\Omega_c h^2$ & 0.1099 \\ 
    $n_s$ & 0.963 \\ 
    $\tau$ & 0.087 \\ 
     $\Delta_R^2$ & $2.41 \times 10^{-9}$ \\ 
   \hline
    $\sigma_0$ for Q & $2.197 \ {\rm mK}$ \\ 
    $\sigma_0$ for V& $3.133 \ {\rm mK}$ \\ 
    $\sigma_0$ for W & $6.538 \ {\rm mK}$\\ 
    $f_{\rm sky}$ & 0.718 \\ 
    \hline
  \end{tabular}
  \caption{Cosmological and noise parameters used in our analysis. The first set is our fiducial cosmology model taken to be
 consistent with WMAP 5-year best-fit cosmology \cite{Komatsu:2008hk}.
The second set of numbers is the normalization parameters related to the instrumental noise in each of the three frequency bands used for the analysis.  $f_{\rm sky}$
is the fraction of sky unmasked by KQ75 mask.}
  \label{tab:params}
\end{table}

\subsection{Unresolved Point Sources}

In addition to the primordial bispectrum, we also account for the
non-Gaussianity generated by unresolved radio point sources.  If the sources
are Poisson distributed, the bispectrum takes a simple from \cite{Komatsu:2001rj} with 
\begin{eqnarray}
B^{\rm PS}_{l_1l_2l_3} &=& I_{l_1l_2l_3} b_{\rm ps},
\end{eqnarray}
where
\begin{equation}
b_{\rm ps}= g^3(x) \int_0^{S_c} S^3 \frac{dn}{dS} \, dS \, ,
\end{equation}
where $dn/dS$ is the number counts of sources and $g(x)$ maps flux density to thermodynamic temperature with $g(x)=c^2(e^x-1)^2/2k_B\nu^2x^2e^x$
with $x=h\nu/k_BT_{\rm CMB} \approx \nu /56.84 {\rm GHz}$. This conversion can be simplified to $g(x) = \mu K/(99.27 {\rm Jy\, sr^{-1}}) (e^x-1)^2/x^4e^x$.
When model fitting to data, we will ignore the exact number counts of the unresolved sources and parameterize the uncertainty with an
overall normalization
\begin{equation}
b_{\rm ps}^i = A_i \times 10^{-25} \, {\rm sr}^2\, ,
\end{equation}
where the index $i$ is for the three bands from WMAP (Q, V, and W) we use here.

Here, we only account for the shot-noise contribution from point sources, similar to the analysis of non-Gaussianity measurements by the WMAP team.
It is likely that unresolved point sources are clustered on the sky, though existing WMAP data with measurements at the two-point function level only lead
to an upper limit on the clustering amplitude of point sources \cite{Serra2}. In future, especially for non-Gaussianity measurement with Planck, it may be necessary
to include the bispectrum generated by clustered point sources.

\subsection{CMB Lensing-Secondary Correlation}

The gravitational lensing effect of the CMB also
generates a bispectrum through correlations of the lensing potential with secondary anisotropies that are generated at late times \cite{CooHu,Goldberg}.

To understand this signal, we note that the
lensed temperature fluctuation in a given direction is the sum of
the primary fluctuation in a different direction plus the secondary
anisotropy
\begin{eqnarray}
T(\bn) &=& T^{\rm P}(\bn + \nabla \Theta) + T^{\rm S}(\bn)  \\
       &\approx&
        \sum_{lm} \Big[ (\almn^{\rm P}+\almn^{\rm S}
                  )\Ylmn(\hat{\bf n})
                +\almn^{\rm P}  \nonumber\\
       &&\times
        \nabla\Theta(\hat{\bf n})\cdot\nabla \Ylmn(\hat{\bf n}) \Big]
\, ,
\nonumber
\end{eqnarray}
or
\begin{eqnarray}
\almn &=& \almn^{\rm P} + \almn^{\rm S}
        + \sum_{l'm'}
        a_{l' m'}^{\rm P}\nonumber\\
&&\times
        \int d \bn \Ylmn{}^* (\bn)
        \nabla\Theta(\hat{\bf n})\cdot
        \nabla Y_{l'}^{m'}(\hat{\bf n})
        \,.
\end{eqnarray}
Utilizing the definition of the bispectrum 
in Eq.~(\ref{eqn:bispectrum}), we obtain
\begin{eqnarray}
\hspace{-0.6cm}
\bi^{\rm lens-sec} &=& \sum_{m_1 m_2 m_3} \wjm
\nonumber \\
&&\times \int d\hat{\bf m} \int d\hat{\bf n}
\Ylm{2}{}^*(\hat{\bf m})
\Ylm{3}{}^*(\hat{\bf n}) C_{l_1}
\nonumber \\
&&\times
\nabla \Ylm{1}{}^*(\hat{\bf m})
\cdot
\langle \nabla\Theta(\hat{\bf m})
T^{\rm S}(\hat{\bf n}) \rangle  + {\rm Perm.}
\end{eqnarray}
where the extra five permutations are with respect to the ordering of
$(l_1,l_2,l_3)$.

Integrating by parts and simplifying further following leads to a
bispectrum of the form:
\begin{eqnarray}
&& \bi^{\rm lens-sec} = -\wj \sqrt{ \frac{(2l_1 +1)(2 l_2+1)(2 l_3+1)}{4 \pi}}  \nonumber \\
&\times&
\left[\frac{l_2(l_2+1)-l_1(l_1+1)-l_3(l_3+1)}{2} C_{l_1}b^{\rm S}_{l_3}+ {\rm Perm.}\right]\,. \nonumber \\
\end{eqnarray}

When calculating the CMB lensing potential-secondary anisotropy cross-correlation $b^{\rm S}_l$ we will 
include both the integrated Sachs-Wolfe (ISW) and the Sunyaev-Zel'dovich (SZ) effects, with the latter modeled
using the halo approach \cite{CooShe,Cooray1,Cooray2}. We will take the sum of the two effects
such that $b_l^S=b_l^{\rm ISW}+b_l^{\rm SZ}$. The cross-correlation between lensing potential and ISW is calculated
in the standard way \cite{Cooray3,Afshordi} for the fiducial $\Lambda$CDM cosmological model, using only the linear theory potential.
For the lensing-SZ correlation, the linear halo model takes into account the SZ profile obtained analytically in Ref.~\cite{KS}
combined with the halo mass function similar to calculations of the SZ angular power spectrum.
When model fitting the data, we will parameterize the overall uncertainty with a parameter $\eta_i$ for each of
the WMAP bands such that $\bi^{\rm lens-sec} \propto \eta_i$.

While lensing modification to CMB bispectrum alone is not expected to make a significant correction to the non-Gaussianity measurement,
analytical calculations of the lensing effect on the CMB bispectrum suggest that the lensing-secondary correlation 
will be the main contamination to a reliable measurement of the primordial non-Gaussianity parameter \cite{Serra,Verde,Hanson,Cooray4}.  
This includes the lensing-ISW effect since SZ can be ``cleaned out'' in multi-frequency data such as those expected from Planck \cite{Cooetal00,Veneziani}.
It is due to this reason that we include the lensing-secondary correlation here. 

\begin{figure}
    \begin{center}
      \includegraphics[scale=0.45]{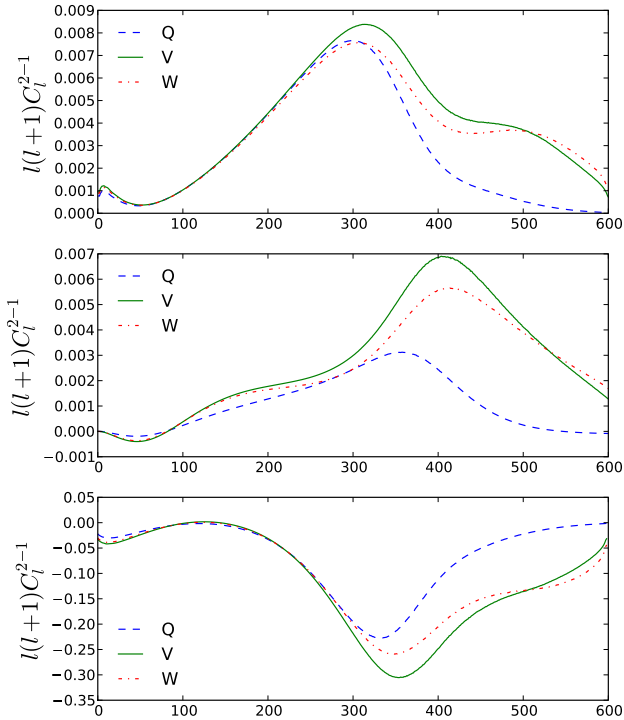} 
   \end{center}
   \vspace{-0.7cm}
   \caption[width=3in]
{Contributions to $C_l^{2-1}$ expected from primordial non-Gaussianity and unresolved point sources.
We show the case with $f_{\rm NL}=1$ for primordial non-Gaussianity (top), shot-noise from unresolved point sources with
$b^i_{\rm ps}=1$ (middle), and lensing-secondary signal  with $\eta_i=1$ (bottom).}
   \label{fig:theoryC21}
\end{figure}

\begin{figure}
    \begin{center}
      \includegraphics[scale=0.45]{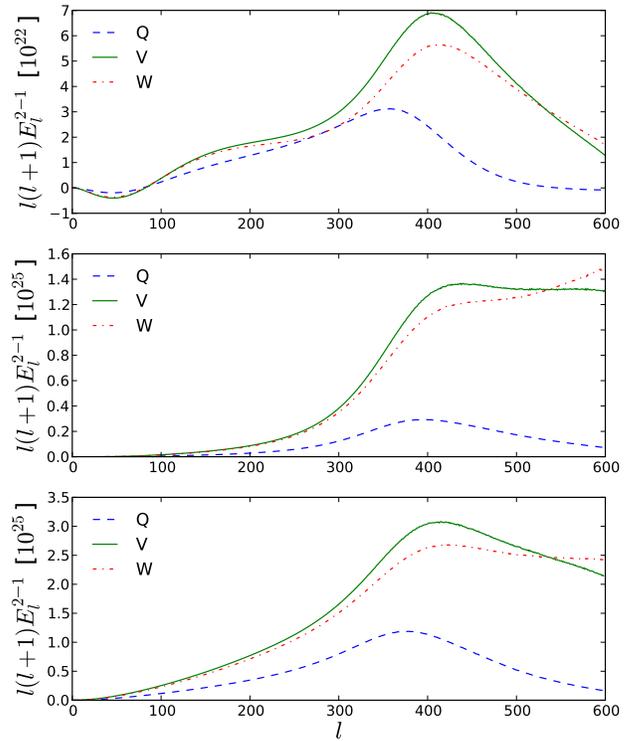} 
   \end{center}
   \vspace{-0.7cm}
   \caption[width=3in]{Contributions to $E_l^{2-1}$ expected from primordial non-Gaussianity and unresolved point sources.
We show the case with $f_{\rm NL}=1$ for primordial non-Gaussianity (top), shot-noise from unresolved point sources with
$b^i_{\rm ps}=1$ (middle), and lensing-secondary signal  with $\eta_i=1$ (bottom). Note the large difference in the y-axis scale from top curve
Involving primordial non-Gaussianity to middle and bottom curves with point sources. As is known, the skewness power spectrum associated with
E maps is more sensitive to shot-noise bispectrum from point sources.}
   \label{fig:theoryE21}
\end{figure}

\section{Estimators of $f_{\rm NL}$}
\label{analysis}

We will now motivate a new estimator for measuring $f_{\rm NL}$. For this we introduce the squared temperature-temperature 
angular power spectrum and discuss its use as a probe of the angular bispectrum.
We motivate a new estimator by revising the original form in Ref.~\cite{Cooray:2001ps}. 

Through the expansion of the temperature
\begin{equation}
T(\bn) = \sum \almn \Ylmn(\bn),
\end{equation}
we can write
\begin{equation}
a_{lm}^2 = \int d\bn T^2(\bn) \Ylmn {}^*(\bn)  \, .
\label{eqn:alm2}
\end{equation}
We emphasize here  that $a_{lm}^2$ denotes the multipole moments of the temperature squared map
and not the square of the multipole moments of the temperature map. 

We can now construct the angular power spectrum of squared temperature and
temperature as
\begin{equation}
 C_l^{2-1} = \frac{1}{2l+1} \sum_m a_{lm}^2 a_{lm}^* \, .
\label{eqn:c2power}
\end{equation}

After some tedious, but straightforward algebra we can write the relation
between the bispectrum of the temperature field and the angular power spectrum
of squared temperature and temperature as
\begin{eqnarray}
&& C_l^{2-1} = \frac{1}{2l+1}\sum_{l_1 l_2} B_{l_1 l_2 l} \\
&\times& 
\left(
                         \begin{array}{ccc}
       l_1 & l_2  & l  \\
         0 & 0 & 0
                         \end{array}
                   \right) 
\sqrt{
        (2 l_1+1) (2 l_2+1) (2l+1) \over 4\pi} \, . \nonumber
\label{eqn:finalform}
\end{eqnarray}
Here, we have made use of the relation
\begin{equation}
\sum_{m_1 m_2} \left(
                         \begin{array}{ccc}
       l_1 & l_2  & l  \\
         m_1 & m_2  & m
                         \end{array}
                   \right)
\left(
                         \begin{array}{ccc}
       l_1 & l_2  & l'  \\
         m_1 & m_2  & m'
                         \end{array}
                   \right) = \frac{\deld_{l l'} \deld_{m m'}}{2 l+1}\, .
\end{equation}
As is clear $C_l^{2-1}$ sums up all triangle configurations of the bispectrum at each of
the side length $l$ of the triangle in multipolar space.  

If a priori known that certain triangular configurations contribute to the
bispectrum significantly one can compute this sum by 
appropriately weighting the multipole coefficients. This is
essentially what can be achieved with the introduction of an appropriate weight
or a window function in equation~(\ref{eqn:alm2}).  Though the analytical expression
for the two-to-one angular power spectrum involves a sum over the two sides of the angular bispectrum, the experimental
measurement is straightforward: one construct the power spectrum by squaring
the temperature field, in real space, and using the Fourier transforms of
squared temperature values and the  temperature field, with any weighting as
necessary.  

This simple form of the skewness power spectrum  has already been used by Szapudi \& Chen
\cite{Szapudi} to constrain $f_{\rm NL}=22 \pm 52$ (1$\sigma$) with WMAP 3-year
data.  The form of the skewness power spectrum as written exactly in
equation~(\ref{eqn:finalform})  is not useful for
a primordial non-Gaussianity measurement.  We describe how to filter data for a
measurement of primordial non-Gaussinity below.

\subsection{Skewness Estimator}

To obtain a more useful form, it is useful to review the form of the skewness
statistic employed by the WMAP team, which is originating from Ref.~\cite{KSW}.  The
skewness statistic makes use of two set of maps of the CMB sky as a function of
the radial distance $r$:
\begin{eqnarray}
A(r,\hat {\mathbf n}) &\equiv& \sum_{lm} Y_{lm}(\hat {\mathbf n}) A_{lm}(r) \\ 
B(r,\hat {\mathbf n}) &\equiv& \sum_{lm} Y_{lm}(\hat {\mathbf n}) B_{lm}(r) \, ,
\end{eqnarray}
where
\begin{eqnarray}
\label{eqn:Alm}
A_{lm}(r) &\equiv& {\alpha_{l}(r) \over {\cal C}_l} b_l a_{lm} \\
\label{eqn:Blm}
B_{lm}(r) &\equiv& {\beta_{l}(r) \over {\cal C}_l} b_l a_{lm} \, .
\end{eqnarray}
Here ${\cal C}_l \equiv C_lb_l^2+N_l$ where $b_l$ are the frequency
dependent beam transfer functions and $N_l$ is the power spectrum from
associated simulated noise maps. We discuss both these quantities later.

In $A$ and $B$ maps weights are such that  they are are constructed from the theoretical CMB power spectrum $C_l$ under the
assumed cosmological model, the experimental beam $b_l$, and the primordial non-Gaussianity projection functions
$\alpha_l(r)$ and $\beta_l(r)$ where $\alpha_l(r)$ and $\beta_l(r)$ are defined in equations~(\ref{eq:alpha}) and (\ref{eq:beta}).

The WMAP team's estimator \cite{KSW} uses an integration in the radial coordinate to obtain the skewness of the product of the $A$ and $B^2$ maps
\begin{equation}
\label{eq:skewness}
S_{\rm AB^2} \equiv \int r^2 dr \int d \hat {\mathbf n}  A(r,\hat {\mathbf n})B^2(r,\hat {\mathbf n}) \, .
\end{equation}
In practice this skewness is corrected by an additional linear term that corrects approximately the
effects of partial sky coverage associated with the mask and non-uniform noise. This term is computed by combining
observed map with simulated maps that are Monte-Carlo averaged (see Appendix A of Ref.~\cite{Komatsu:2008hk}). 

As is clear from above $S_3$ involves a complete compression of data to a single number.
While in principle different sources of non-Gaussianities contribute to $S_3$ with a single number alone it is impossible to
separate out the primordial value from the non-Gaussianities generated by secondary anisotropies and other foregrounds. To some extent the separation
is aided by a different set of maps that are weighted differently than the case of $A$ and $B$ maps.

A map optimized for the non-Gaussianity of the form generated by shot-noise from point sources is the $E$ map:
\begin{eqnarray}
\label{emap}
E(\hat {\mathbf n}) &\equiv& \sum_{lm} Y_{lm}(\hat {\mathbf n}) E_{lm}(r) \, ,
\end{eqnarray}
where
\begin{eqnarray}
\label{eqn:elm}
E_{lm}(r) &\equiv& {b_l \over {\cal C}_l} a_{lm} \, .
\end{eqnarray}
Similar to $S_{\rm AB^2}$, one can also compute a skewness associated with E maps by taking $S_{\rm E^3} =
\int d \hat {\mathbf n}  E^3(\hat {\mathbf n})$. WMAP team used the latter to constrain the normalization of the point source Poisson
term with $b_{\rm ps}$.

\begin{figure}
    \hspace{-3cm}
    \begin{center}
      \includegraphics[scale=0.5]{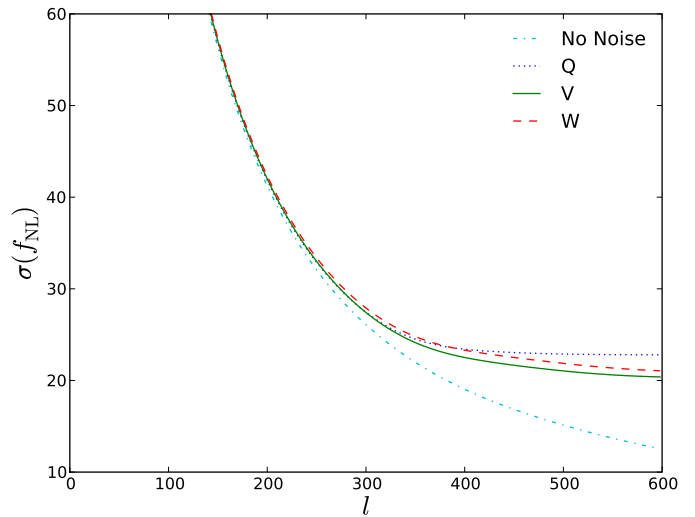} 
   \end{center}
   \vspace{-0.7cm}
   \caption[width=3in]{Expected error for $f_{\rm NL}$ calculated based on the Fisher matrix approach for each of the three noise curves for
the WMAP in Q, V, and W-bands and with $f_{\rm sky}=0.718$ when using KQ75 mask. The Cramer-Rao bound ranges from about $\sim 21$ in V-band to
$\sim$ 23 in Q-band. This estimate assumes that only the primordial non-Gaussianity signal is present in the bispectrum and ignores the
degeneracies between primordial non-Gaussianity and other parameters, such as those related to unresolved point sources.}
   \label{fig:fisher}
\end{figure}

\subsection{Revised Skewness Power Spectrum}

In order to revise the previously discussed skewness power spectrum, instead of simply integrating over the
$A$ and $B^2$ maps, we extract the multipole moments of the $B^2$ map and the product $AB$ maps
\begin{eqnarray}
\left(B^2\right)_{lm}(r) &\equiv& \int d\hat {\mathbf n} B^2(r,\hat {\mathbf n})Y_{lm}(\hat {\mathbf n})  \, \nonumber \\
\left(AB\right)_{lm}(r) &\equiv&  \int d\hat {\mathbf n} A(r,\hat {\mathbf n})B(r,\hat {\mathbf n})Y_{lm}(\hat {\mathbf n})  \, .
\end{eqnarray}
These two multipole moments then allow us to write the new skewness power spectrum  appropriately weighted in the
same manner as the previous skewness estimator:
\begin{eqnarray}
\label{eq:C21}
C_l^{2-1} &\equiv& (C_l^{A,B^2} + 2 C_l^{AB,B}) \\
C_l^{A,B^2}&\equiv& \frac{1}{2l+1} \int r^2 dr \left[ \sum_m {\rm Real}\left \{ A_{lm}(r) \left(B^2\right)_{lm}(r) \right \} \right] \nonumber \\
C_l^{B,AB} &\equiv& \frac{1}{2l+1} \int r^2 dr \left[ \sum_m {\rm Real} \left \{ B_{lm}(r) (AB)_{lm}(r) \right \} \right] \, . \nonumber 
\end{eqnarray}

To see how $C_l^{2-1}$ probes the primordial bispectrum, we can write the multipole moments of the squared B map as
\begin{eqnarray}
\label{eq:B_sq}
&&\left(B^2\right)_{lm}(r)  =\\
&& 
\sum_{l'm'} \sum_{l''m''} {\beta_{l'}(r) \over {\cal C}_{l'}}{\beta_{l''}(r) \over {\cal C}_{l''}}
 \sqrt {(2l+1)(2l'+1)(2l''+1) \over 4\pi} \nonumber \\
&\times&\left ( \begin{array}{ c c c }
     l & l' & l'' \\
     0 & 0 & 0
  \end{array} \right)
\left ( \begin{array}{ c c c }
     l & l' & l'' \nonumber \\
     m & m' & m''
  \end{array} \right)a'_{l'm'} a'_{l''m''} \, , \nonumber 
\end{eqnarray}
where $a'_{lm}$ are the beam times the observed multipole moments $(b_l a_{lm})$. Note that the observed multipole moments
relate to theory moments via another beam factor.

Similarly, the multipole moments of the $(AB)$ product map is
\begin{eqnarray}
\label{eq:AB}
&&\left(AB\right)_{lm}(r)  = \\
&& \sum_{l'm'} \sum_{l''m''} {\alpha_{l'}(r) \over {\cal C}_{l'}}{\beta_{l''}(r) \over {\cal C}_{l''}}
 \sqrt {(2l+1)(2l'+1)(2l''+1) \over 4\pi} \nonumber \\
&\times&\left ( \begin{array}{ c c c }
     l & l' & l'' \\
     0 & 0 & 0
  \end{array} \right)
\left ( \begin{array}{ c c c }
     l & l' & l'' \\
     m & m' & m''
  \end{array} \right)a'_{l'm'} a'_{l''m''} \, . \nonumber 
\end{eqnarray}

The $C_l^{A,B^2}$ power spectrum is simply then
\begin{eqnarray}
&&C_l^{A,B^2} = \frac{1}{2l+1}\int r^2 dr \sum_m \sum_{l'm'} \sum_{l''m''} \\
&\times& {\beta_{l'}(r) \over {\cal C}_{l'}}{\beta_{l''}(r) \over {\cal C}_{l''}}
  {\alpha_{l}(r) \over {\cal C}_{l}}
 \sqrt {(2l+1)(2l'+1)(2l''+1) \over 4\pi} \nonumber \\
&\times&\left ( \begin{array}{ c c c }
     l & l' & l'' \\
     0 & 0 & 0
  \end{array} \right)
\left ( \begin{array}{ c c c }
     l & l' & l'' \\
     m & m' & m''
  \end{array} \right)a'_{lm} a'_{l'm'} a'_{l''m''} \, . \nonumber
\end{eqnarray}
Using the definition of the angular bispectrum, we can simplify to obtain
\begin{eqnarray}
&&C_l^{A,B^2} =  \frac{1}{2l+1}\int r^2 dr \sum_{l'l''}  \\
&\times& {\beta_{l'}(r) \over {\cal C}_{l'}}{\beta_{l''}(r) \over {\cal C}_{l''}}
  {\alpha_{l}(r) \over {\cal C}_{l}}
 \sqrt {(2l+1)(2l'+1)(2l''+1) \over 4\pi} \nonumber \\
&\times&\left ( \begin{array}{ c c c }
     l & l' & l'' \\
     0 & 0 & 0
  \end{array} \right) \hat{B'}_{ll'l''} b_l b_l' b_l'' \, ,
\end{eqnarray}
where $\hat{B'}_{ll'l''}$ is the bispectrum estimated from data under beam smoothing. It relates to the theory bispectrum $B_{ll'l''}$ as
$\hat{B}'_{ll'l''}=B_{ll'l''}b_l b_l' b_l''$.

We can similarly simplify the term for $C_l^{AB,B}$ and putting the two terms together, we find that
the total is
\begin{eqnarray}
\label{eq:C21final}
C_l^{2-1} &\equiv& (C_l^{A,B^2} + 2 C_l^{AB,B}) \\
&=& {1 \over (2l+1)} \left[\sum_{l'l''} \left \{ B^{NG,f_{\rm NL}=1}_{ll'l''} \hat B'_{ll'l''} b_l b_l' b_l'' \over {\cal C}_l {\cal C}_{l'} {\cal C}_{l''} \right \} \right] \, . \nonumber
\end{eqnarray}

If we assume that the observed bispectrum is simply that of the primordial non-Gaussianity then $\hat B_{ll'l''} = \hat{f}_{\rm NL}B^{NG}_{ll'l''}$
and we can write an estimator for $f_{\rm NL}$ as
\begin{equation}
\hat{f}_{NL} = (2l+1)C_l^{2-1}/F_{{\rm NG},{\rm NG}}(l) \, ,
\end{equation}
where $F_{{\rm NG},{\rm NG}}(l)$ is simply the Fisher matrix element for the primordial bispectrum with $f_{\rm NL}=1$:
\begin{equation}
F_{i,j}(l) = \sum_{ll''} \left \{ B^i_{ll'l''} B^j_{ll'l''} \over {\cal C'}_l {\cal C'}_{l'} {\cal C'}_{l''} \right \} \, ,
\end{equation}
where now we have redefined noise to be such that ${\cal C'}_l=C_l+N_l/b_l^2$ as the bispectra are no longer beam smoothed.

In reality $C_l^{2-1}$ includes contributions from secondary anisotropies and foregrounds. Here, 
we include the non-Gaussianities generated by point sources and the lensing-secondary correlation. 
Thus, we write
\begin{equation}
\label{eq:cl21}
(2l+1)\hat{C}_l^{2-1} = \hat{f}_{NL} F_{{\rm NG},{\rm NG}}(l)   + \hat{A} F_{{\rm NG},{\rm PS}} + \hat{\eta} F_{{\rm NG},{\rm len-sec}} \, ,
\end{equation}
and consider a joint estimation of the three unknown parameters.

To help break degeneracies between the three parameters, we also estimate the skewness power spectrum of the E map defined in
equation~(\ref{emap}) as
\begin{equation}
\label{eq:clee2}
C_l^{E,E^2} \equiv \frac{1}{2l+1} \left[ \sum_m {\rm Real}\left \{ E_{lm} \left(E^2\right)_{lm} \right \} \right]  \, .
\end{equation}
Similar to our derivation above one can simplify the multipole moments of the $(E^2)_{lm}$ to show that this probes
\begin{eqnarray}
\label{eq:E21}
E_l^{2-1} &\equiv& C_l^{E,E^2} \\
&=& {1 \over (2l+1)} \left[\sum_{l'l''} \left \{ B^{PS,b_{\rm ps}=1}_{ll'l''} \hat B'_{ll'l''} b_l b_l' b_l'' \over {\cal C}_l {\cal C}_{l'} {\cal C}_{l''} \right \} \right] \, . \nonumber
\end{eqnarray}

Thus, we write
\begin{equation}
\label{eq:el21}
(2l+1)\hat{E}_l^{2-1} = \hat{f}_{NL} F_{{\rm PS},{\rm NG}}(l)   + \hat{A} F_{{\rm PS},{\rm SN}} + \hat{\eta} F_{{\rm PS},{\rm lens-sec}} \, .
\end{equation}

The two equations~(\ref{eq:cl21}) and (\ref{eq:el21}) will form the main set of equations that we will solve with our measurements. 
While we have not explicitly stated so far, these two quantities will be measured in 3 WMAP frequency channels making use of Q, V, and W-band data.
We allow for frequency dependence in $A$ and $\eta$, but assume $f_{\rm NL}$ is the same independent of the frequency in all three channels.

\begin{figure*}
    \begin{center}
      \includegraphics[scale=0.4]{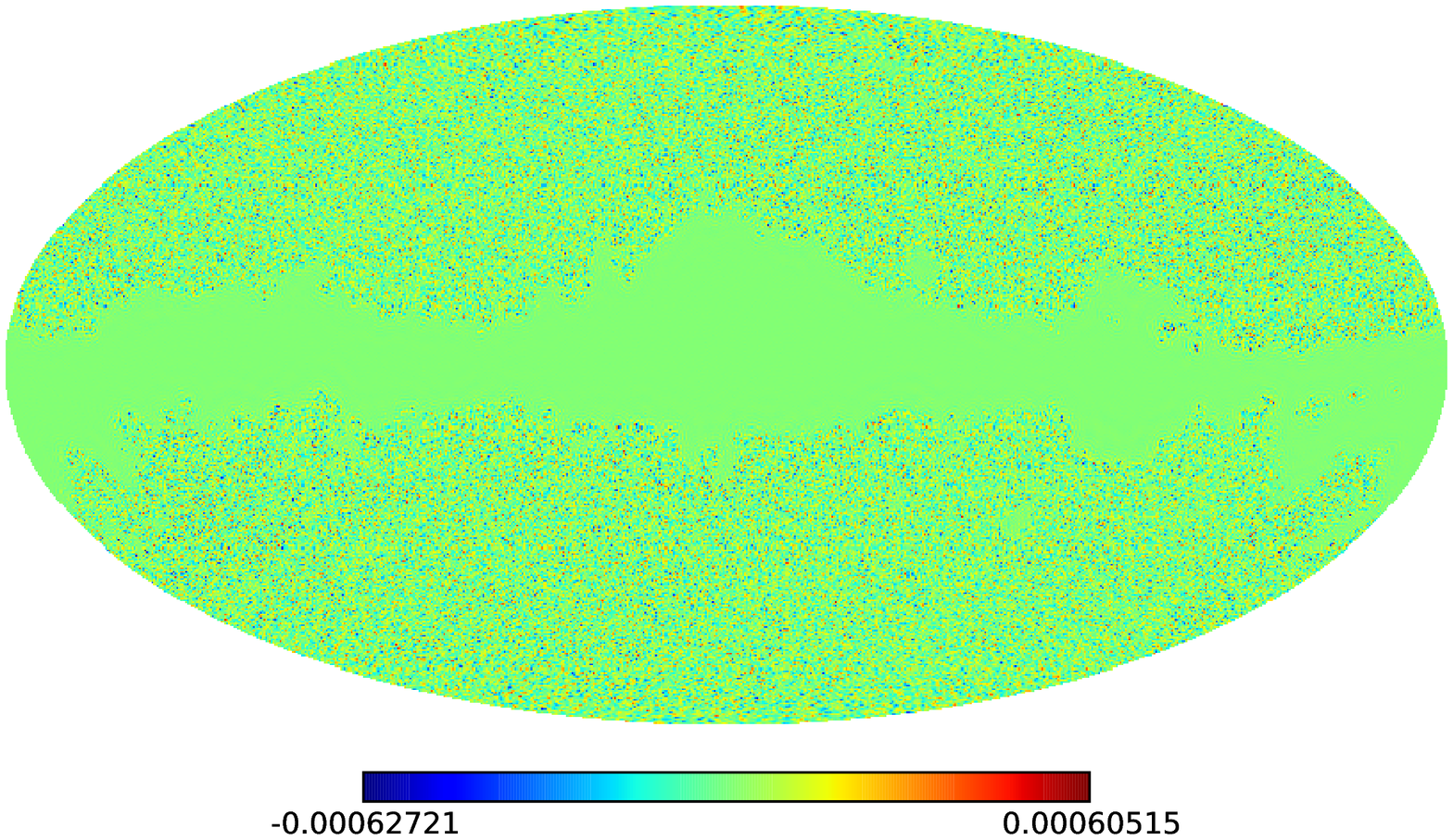} 
      \includegraphics[scale=0.4]{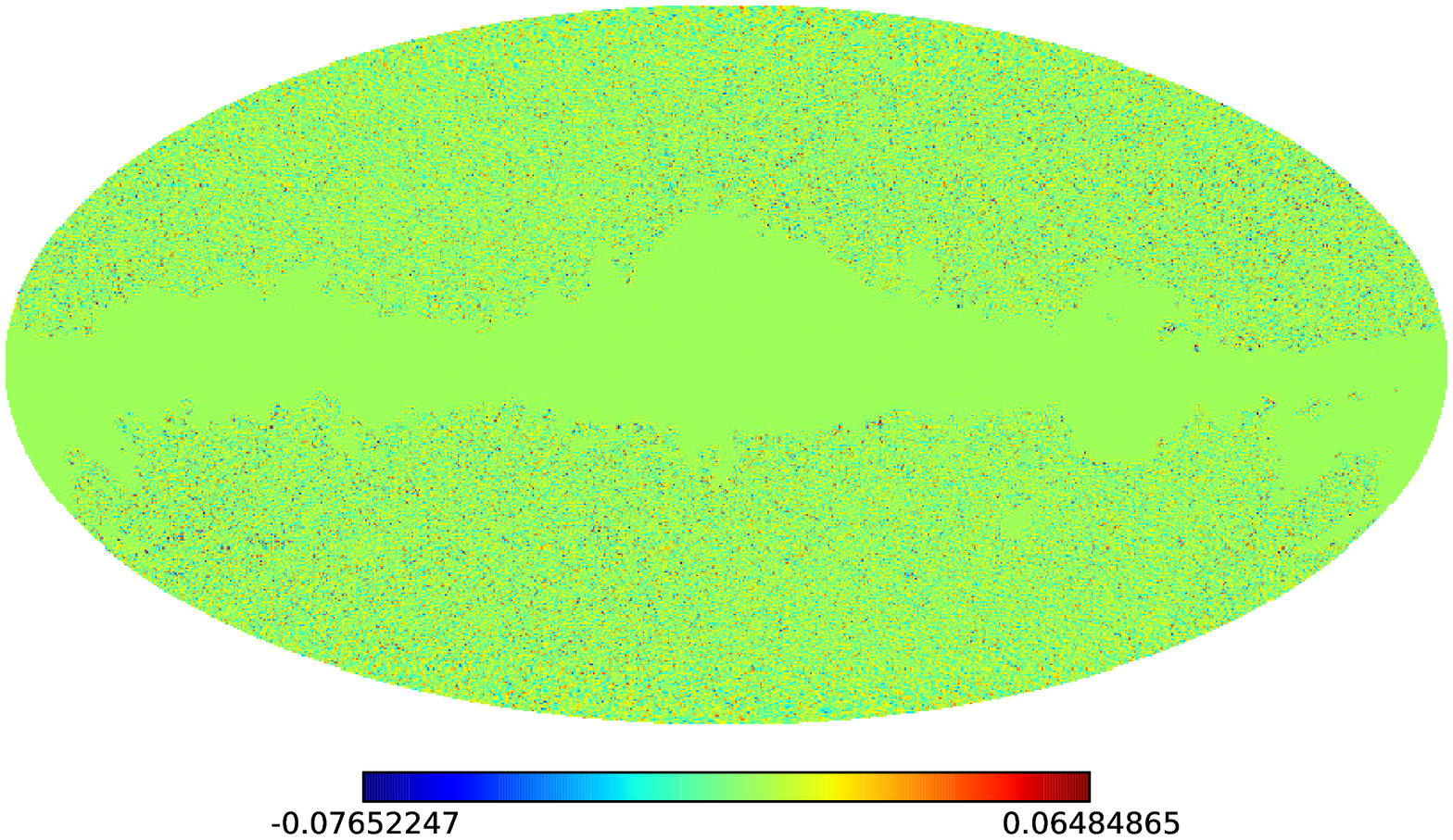} 
      \includegraphics[scale=0.4]{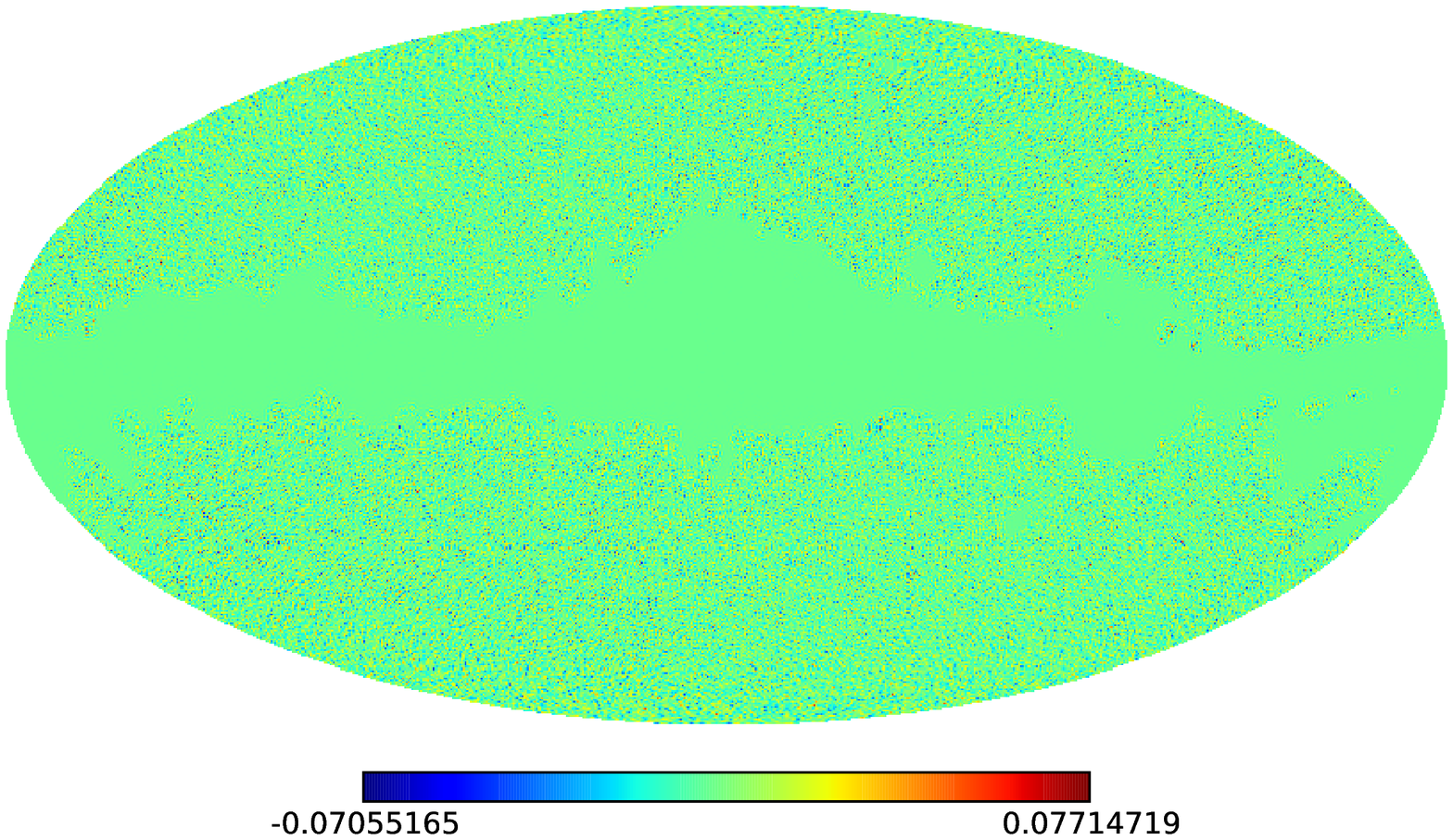} 
        \includegraphics[scale=0.4]{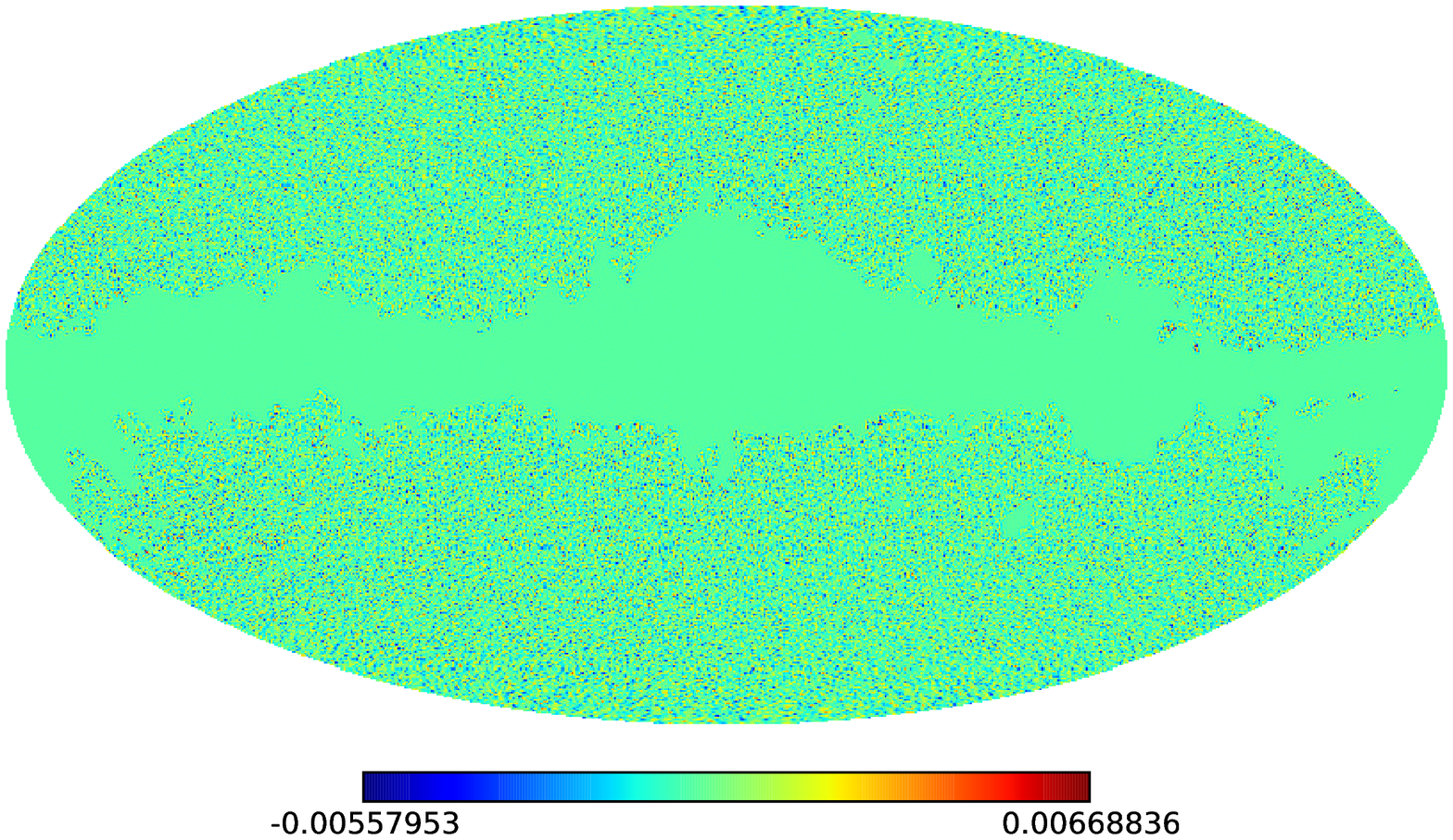} 
   \end{center}
   \caption[width=3in]{A maps for V frequency band.  From upper left hand corner moving clockwise:  $\tau = 0.04, 0.4, 1.0, 1.8$}
   \label{fig:Amaps}
\end{figure*}

\begin{figure*}
    \begin{center}
      \includegraphics[scale=0.4]{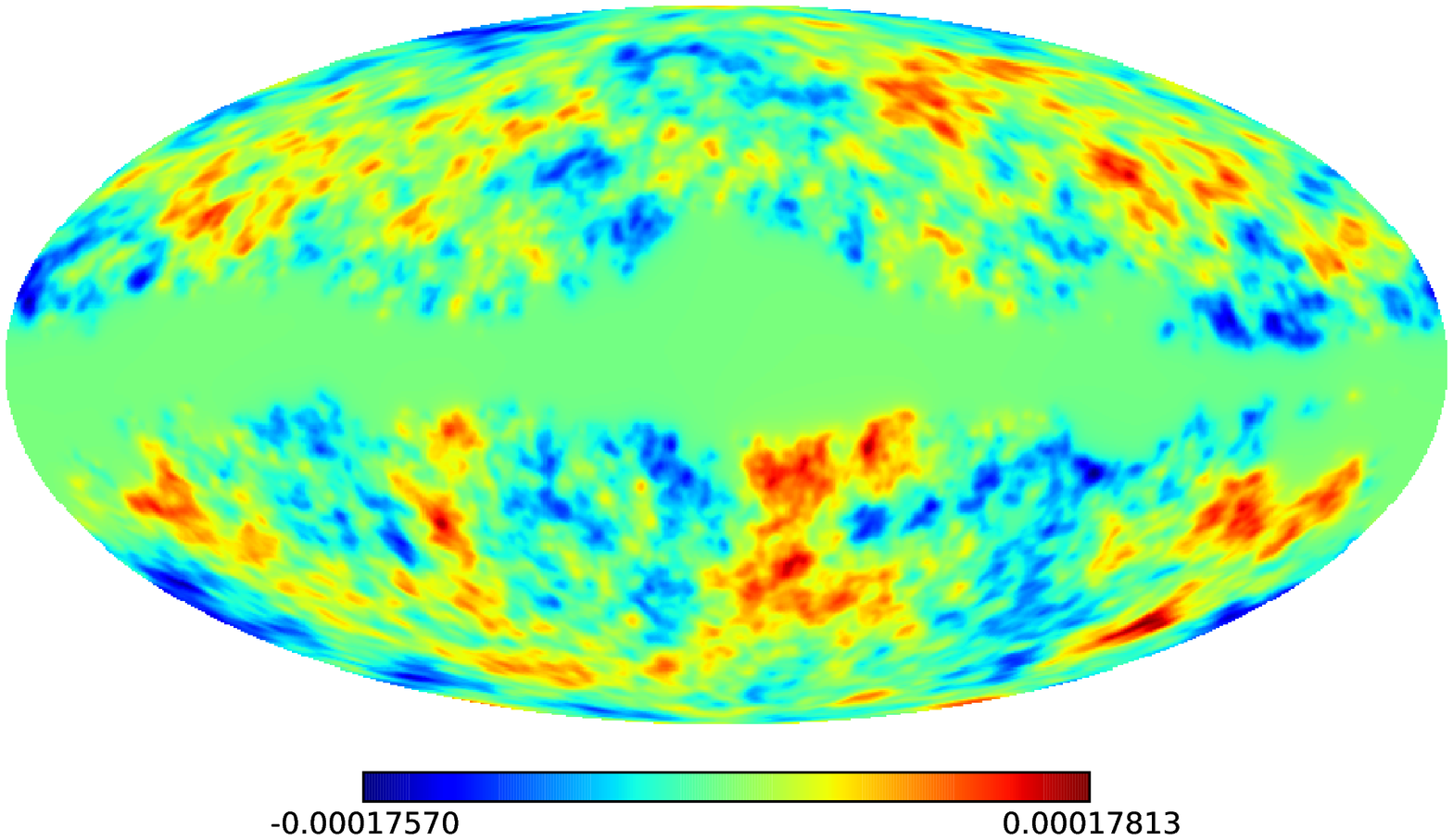} 
      \includegraphics[scale=0.4]{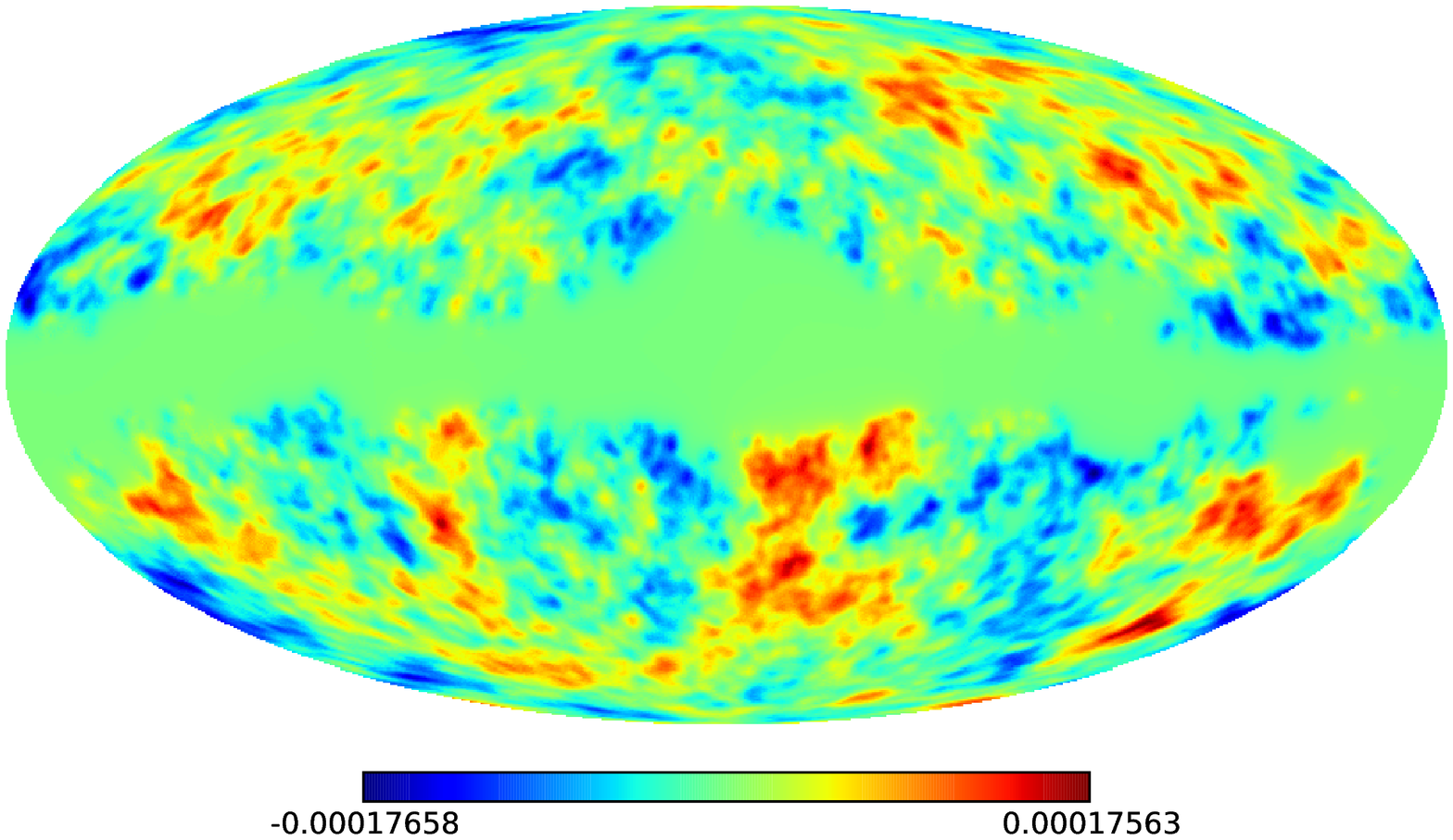} 
      \includegraphics[scale=0.4]{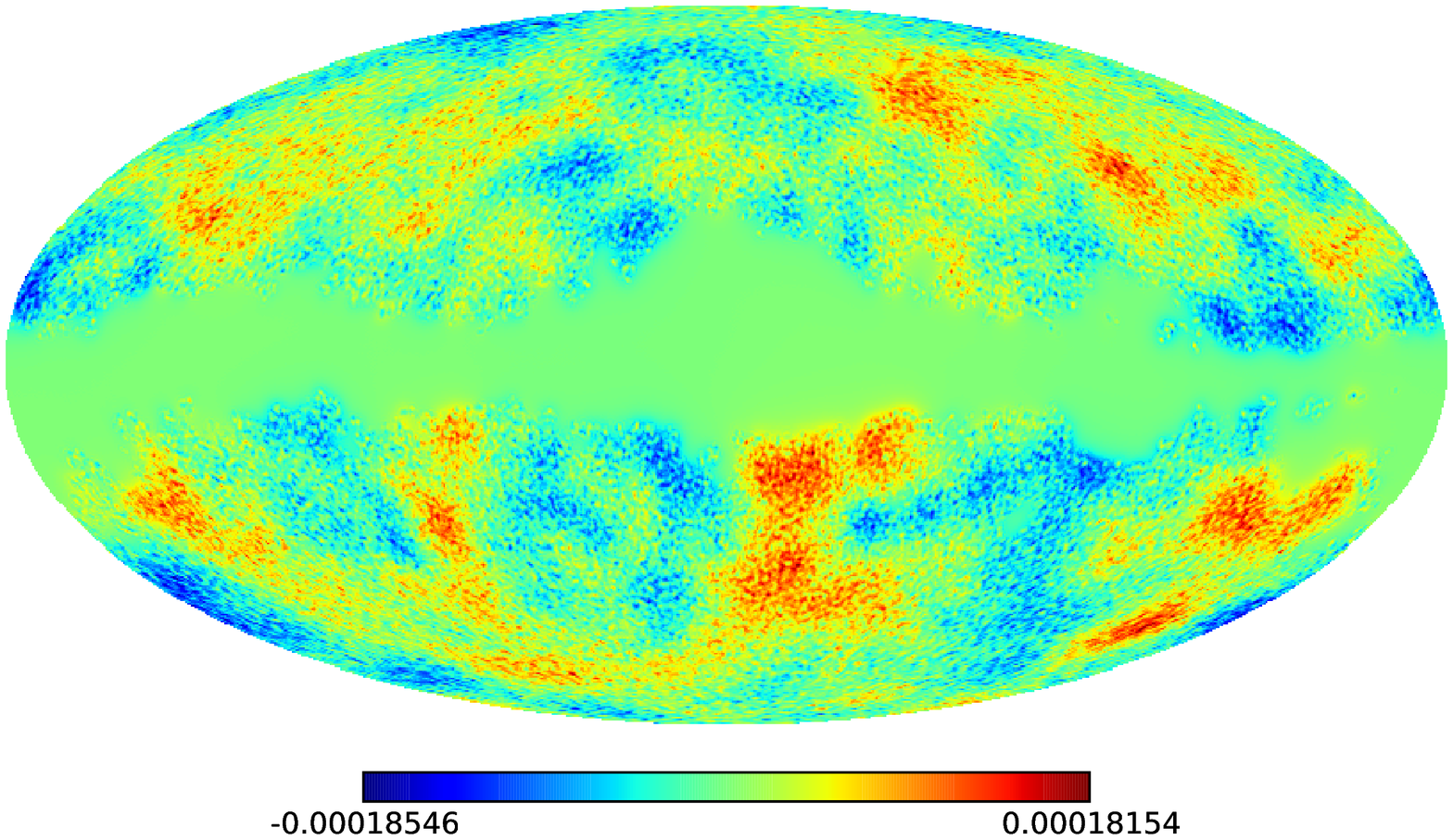} 
        \includegraphics[scale=0.4]{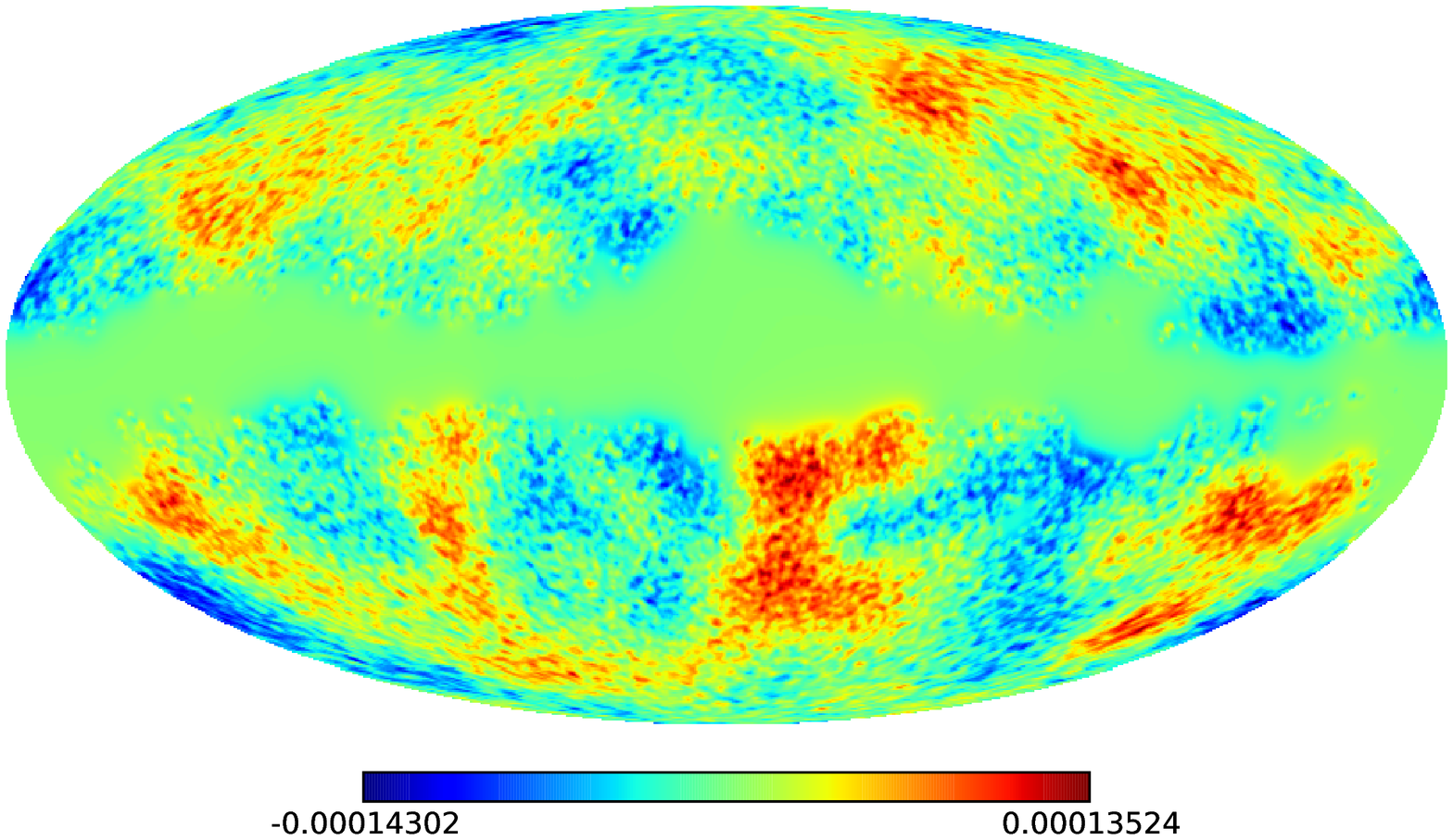} 
   \end{center}
   \caption[width=3in]{B maps for V frequency band.  From upper left hand corner moving clockwise:  $\tau = 0.04, 0.4, 1.0, 1.8$}
   \label{fig:Bmaps}
\end{figure*}

\begin{figure*}
    \begin{center}
      \includegraphics[scale=0.4]{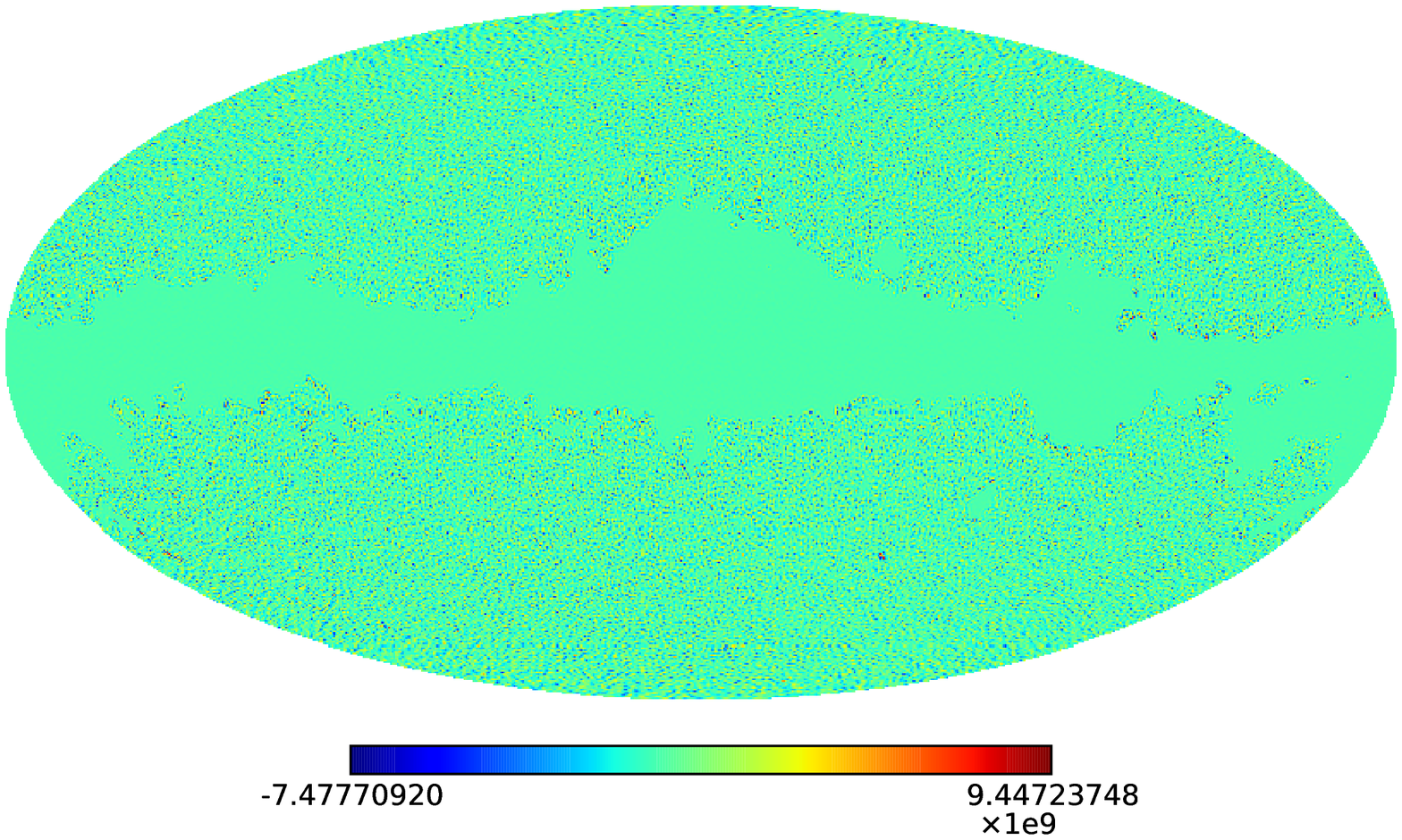} 
      \includegraphics[scale=0.4]{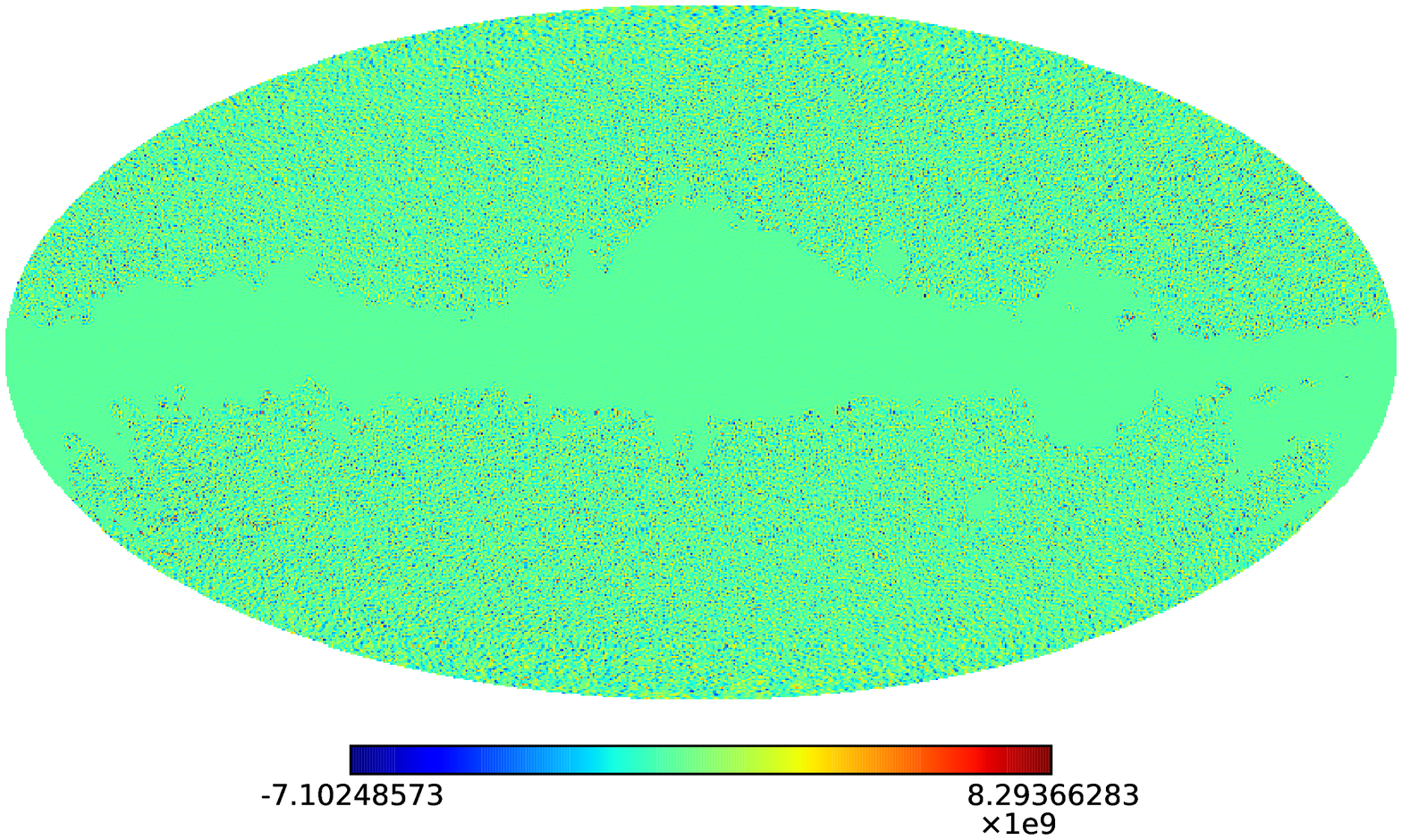} 
      \includegraphics[scale=0.4]{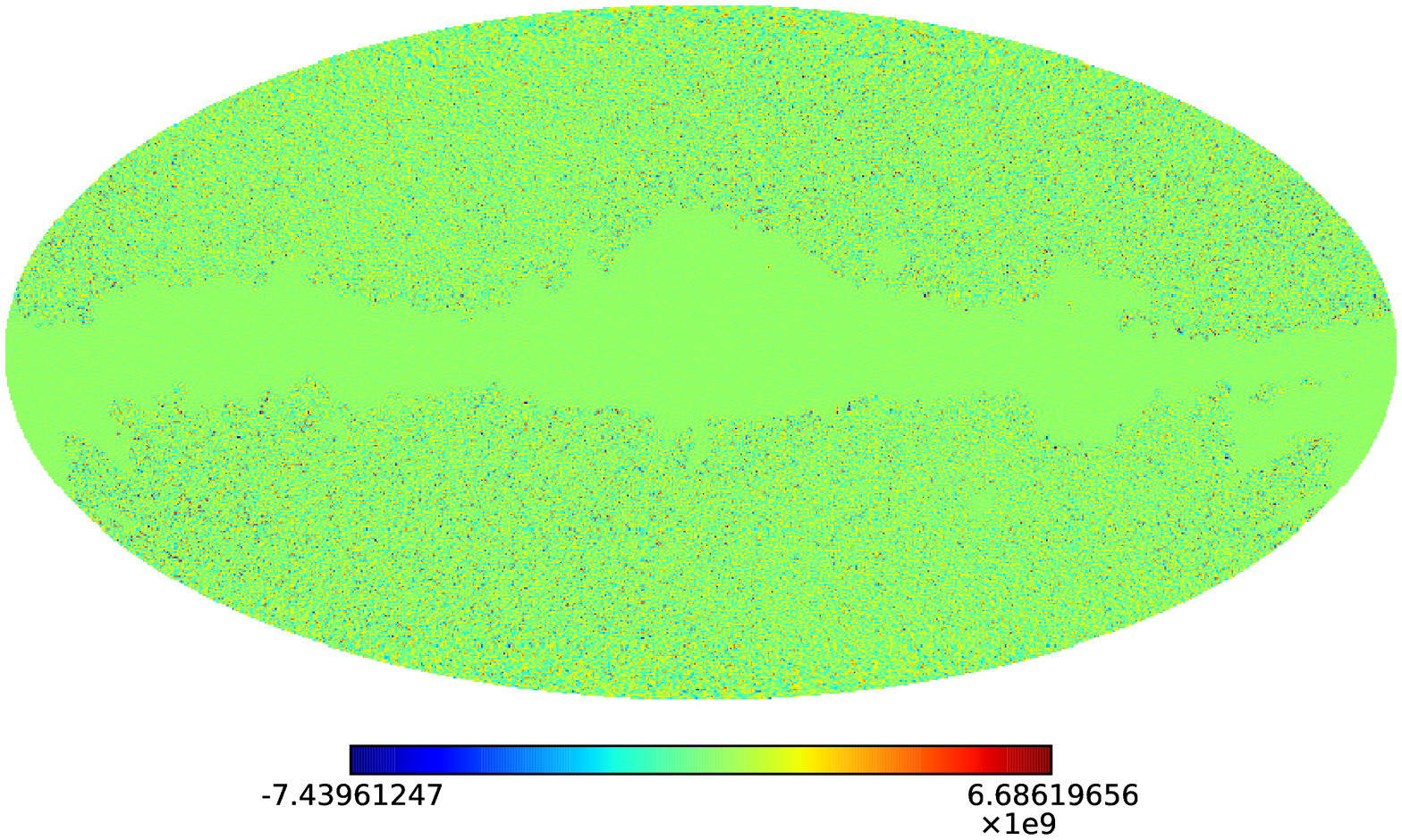} 
   \end{center}
   \caption[width=3in]{E maps for Q, V and W frequencies.}
   \label{fig:Emaps}
\end{figure*}

\subsection{Approximate corrections for partial sky}

Before we move onto discuss data analysis and our simulations to compute the covariances, we note that we also make a correction to both
$C_l^{2-1}$ and $E_l^{2-1}$ to account for partial sky coverage and inhomogeneous noise. This is done in an approximate manner by making use of the equivalent form
of the linear terms of the skewness statistic in the language of our skewness power spectrum. For the case of $C_l^{2-1}$ estimator the correction
is derived in Ref.~\cite{Munshi:2009ik}:
\begin{eqnarray}
\label{eq:C21tot}
C_l^{2-1} = {1 \over f_{sky}}\left \{ {C_l^{A,B^2} - 
2C_l^{\langle A,B \rangle B} - C_l^{A, \langle B^2 \rangle}} \right \} +
\nonumber \\
{2 \over f_{sky}}\left \{ {C_l^{AB,B} - 
C_l^{\langle AB \rangle, B} -C_l^{B\langle A, B\rangle} - C_l^{A\langle B,B \rangle}} \right \}
\end{eqnarray}
where $f_{sky}$ is the sky fraction observed. The new terms are
defined as, for example,
\begin{equation}
\label{eq:clbab}
C_l^{B\langle A, B\rangle}(r) = {1 \over N (2l + 1)} \sum_i \sum_m \left
\{ (B^D A^S)^i_{lm}(r)
(B^S)_{lm}^i(r) \right \},
\end{equation}
where $i$ runs over a set of $N$ simulations and $(B^D A^S)^i_{lm}$ are the
coefficients of spherical harmonics for the map produced by multiplying
the $i^{\rm th}$ simulated $A$ map with the $B$ map derived from raw
data. 

Similarly, for $E_l^{2-1}$ we find
\begin{eqnarray}
\label{eq:E21tot}
E_l^{2-1} = {1 \over f_{sky}}\left \{ C_l^{E,E^2} - C_l^{E, \langle E^2 \rangle} - 2C_l^{\langle E,E \rangle E} \right \} \, ,
\end{eqnarray}
where terms such as $C_l^{E, \langle E^2 \rangle}$ can be written similar to equation~(\ref{eq:clbab}) above
with the replacement of E maps instead of A and B maps.

\subsection{Theoretical expectation}

In Figure~\ref{fig:theoryC21}  and Figure~\ref{fig:theoryE21} we show the theoretical expectations for
$C_{l}^{2-1}$ and  $E_{l}^{2-1}$, respectively. We plot these for the Q, V and W band by making use of the
beam functions $b_l$ and noise power spectrum estimate $N_l$ that are described in Section~\ref{noise}.
Here, we show the cases of primordial non-Gaussianity with $f_{\rm NL}=1$, point sources with $A_i=1$
and lensing-secondary cross-correlation with $b_l^{S}$ calculated for the sum of ISW and SZ
effects with $\eta_i=1$. 

As is clear from Fig.~2, the primordial non-Gaussianity signal is expected to be degenerate
with foreground non-Gaussianities. The shape of $C_l^{2-1}$ alone is not enough to clearly separate
primordial non-Gaussianity signal from point source and lensing non-Gaussianities. Fortunately, the separation is aided
when $C_l^{2-1}$ is combined with $E_l^{2-1}$. As is clear from Fig.~3 (espcially note the difference in the y-axis
range for the top and middle plots), this latter power spectrum allows a better determination of the
point sources. In practice, we perform a combined analysis of both spectra, including the confusion from secondary bispectra,
when model fitting to quantities, $f_{\rm NL}$,$A_i$ and $\eta_i$. To compare with previous results on the literature related to the
non-Gaussianity parameter with WMAP data using the effects of point sources only, we also consider the case where lensing is ignored in the analysis.

In Fig.~4, we include a plot of the $F^{-1/2}_{{\rm NG},{\rm NG}}(l)$, showing the expected error $f_{\rm NL}$
as a function of the multipole for Q, V, and W bands. Out to $l_{\rm max}$ of 600 and with $f_{\rm sky}=0.718$, 
the Carmer-Rao bound is at the level of $\sim$ 21 with V-band to $\sim$ 23 with Q-band. This assumes that
bispectrum only contains primordial non-Gaussianity, but the degeneracy between secondary non-Gaussian signals
and primordial non-Gaussianity is expected to increase the optimal error at some level more than this bound.
Also, to saturate the Cramer-Rao bound an optimal estimator that accounts for the mode-mode correlations
associated with the partial sky and the mask will become necessary \cite{Smith:2009jr}. Our estimator only accounts for the cut-sky approximately
making use of the linear terms. We also weight each multipole coefficient with $(C_lb_l^2+N_l)^{-1}$, as in the case of Gaussian statistics appropriate for the whole sky.
While this approach is not different from that of the WMAP team's \cite{Komatsu:2008hk}, in
an upcoming paper we hope to return to the issue of an exact calculation implementing the full covariance
for the two-to-one skewness power spectrum.

\begin{figure}
    \begin{center}
      \includegraphics[scale=0.45]{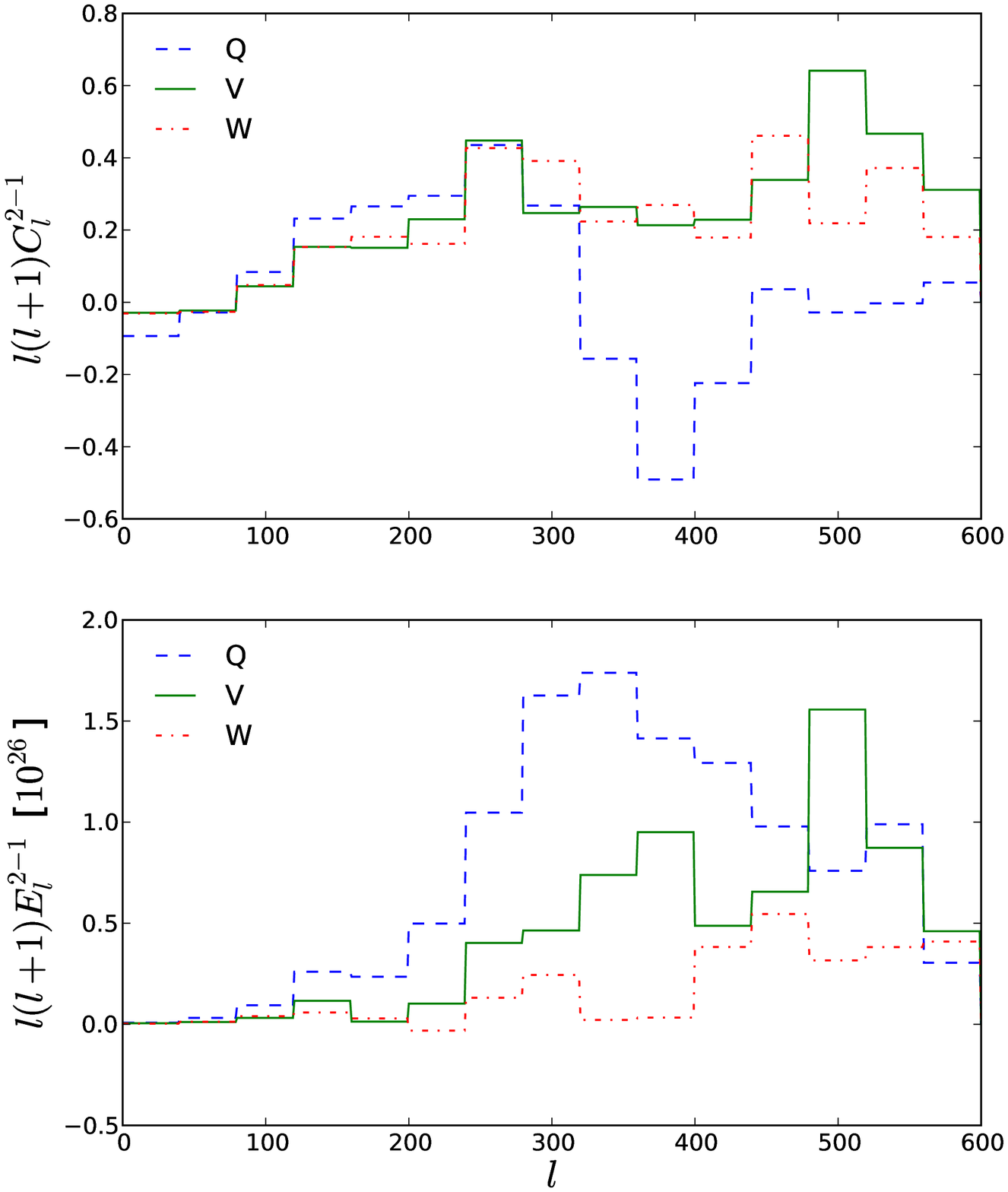} 
   \end{center}
   \vspace{-0.7cm}
   \caption[width=3in]{$C_{l}^{2-1}$ (top) $E_{l}^{2-1}$ (bottom) for Q, V and W with the measured spectra binned with a width of $\delta l = 40$. }
   \label{fig:c21qvw}
\end{figure}

\section{Data Analysis}
\label{sec:data}

We first discuss our data analysis procedure and then how we computed the covariance through simulations.

\subsection{Measurement of $C_{l}^{2-1}$ and $E_l^{2-1}$}

To extract $C_{l}^{2-1}$ and $E_l^{2-1}$ from data we use the raw WMAP 5-year Stokes-I
sky maps for the Q, V and W frequency bands as available from the public lambda website\footnote{http://lambda.gfsc.nasa.gov}.  
We use Healpix\footnote{For more information see http://healpix.jpl.nasa.gov} \cite{Gorski:2004by} to analyze the maps.  Specifically, starting from
the fits files of raw maps we use {\it anafast}, masking with the $KQ75$ mask and without an
iteration scheme, to generate multipole coefficients ($a_{lm}$s) for each frequency map out to $l_{\rm max} =
600$. We will refer to these multipole moments hereafter as as $a_{lm}^D$.  With these definitions
we use equations~(\ref{eqn:Alm}), (\ref{eqn:Blm}), and (\ref{eqn:elm}) to generate $A_{lm}$,
$B_{lm}$, and $E_{lm}$ by substituting $a_{lm}^D$ in place of $a_{lm}$.

Our recipe for obtaining $C_{l}^{2-1}$ and $E_l^{2-1}$ is:
\begin{enumerate}
\item Use Healpix and the $KQ75$ mask to generate $a_{lm}^{D}$ from the WMAP 5-year Stokes-I Sky Maps for the Q, V and W frequency bands.
\item Obtain $A(r,\hat {\mathbf n})$, $B(r,\hat {\mathbf n})$ and $E(\hat {\mathbf n})$ from equations~(\ref{eqn:Alm}), (\ref{eqn:Blm}), and (\ref{eqn:elm}) by
 using $\alpha (r)$ and $\beta (r)$ from equation~(\ref{eq:alpha}) and (\ref{eq:beta}), respectively, with the replacement of $a_{lm}$ with $a_{lm}^{D}$.  
Figures~\ref{fig:Amaps}, \ref{fig:Bmaps} and \ref{fig:Emaps} show the resulting maps.
\item Calculate $C_{l}^{A,B^{2}}$, $C_{l}^{B,AB}$ and the linear terms from equations~(\ref{eq:C21}) and (\ref{eq:C21tot}), respectively, with latter
using equations of the form (\ref{eq:clbab}). Repeat the same to obtain $C_l^{E,E^2}$ with E maps as defined in equation~(\ref{eq:clee2}) 
and the corresponding equations for linear terms
in equation~(\ref{eq:E21tot}).  These correction associated with partial sky coverage involves the use of simulated maps described below.
We integrate over $r$ from  $\tau=0.004$ to 2 with 500 steps. (see Fig.~1).
\item Use the estimated $C_{l}^{2-1}$ and $E_{l}^{2-1}$ with WMAP Q, V, and W maps for our parameter estimate analysis (see below). 
\item Compute analytically $F_{ij}$ terms with $l_{\rm max} =600$ in each of the summations of $l_1$ and $l_2$ and making use of
the noise and beam spectra for WMAP (see below).
\end{enumerate}

In Figure~\ref{fig:c21qvw} we see the $C_{l}^{2-1}$  for each WMAP frequency
band plotted as a function of $l$.
These plots were generated by binning the estimators with $\delta l$ of
40 and plotting the midpoint of each bin.  The V and W estimators have
roughly the same shape and are mostly positive.  The Q  estimator is
noticeably different, dropping negative when $l > 300$.

Furthermore, in Figure~\ref{fig:c21qvw} we see $E_{l}^{2-1}$  for each WMAP frequency
band plotted as a function of $l$.   Like the estimators mentioned above, these were similarly binned in bins of size $\delta l = 40$.  

\begin{figure}
    \begin{center}
      \includegraphics[scale=0.45]{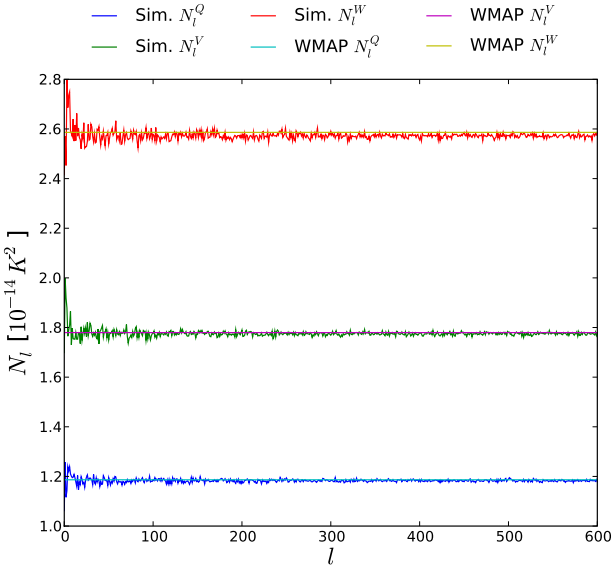} 
   \end{center}
   \vspace{-0.7cm}
   \caption[width=3in]{The noise power spectrum from our masked simulated noise
maps for the Q, V and W frequency bands compared with the analytical
results from the WMAP 5-year team.}
   \label{fig:Nl}
\end{figure}

\begin{figure}
    \begin{center}
      \includegraphics[scale=0.45]{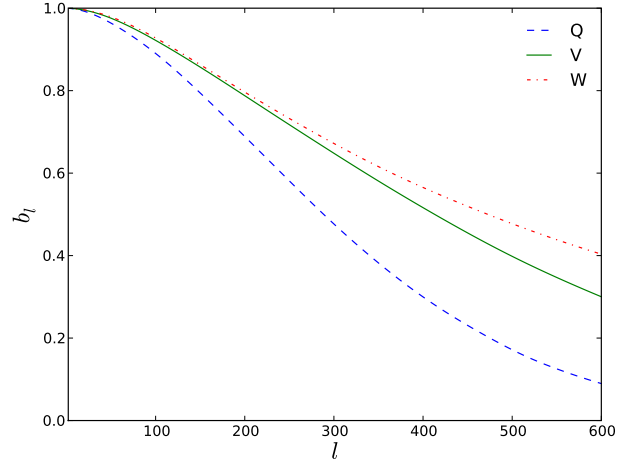} 
   \end{center}
   \vspace{-0.7cm}
   \caption[width=3in]{The beam transfer functions, $b_l$, used in our analysis for each frequency Q, V and W.}
   \label{fig:bl}
\end{figure}

Lastly, in Figures \ref{fig:tot} and \ref{fig:totE} we see the contributions to $C_{l}^{2-1}$ and $E_{l}^{2-1}$ from each term in equations \ref{eq:C21tot} and \ref{eq:E21tot} respectively.  The linear terms are not very significant compared to the other terms.  Nevertheless, they are still considered in this analysis.

\subsection{Simulations of $A$, $B$, and $E$ maps}
\label{sec:simul}

In order to do proper statistics for $C_{l}^{2-1}$ and $E_l^{2-1}$, 
we create 250 simulated maps at each WMAP frequency band. To do so we first produce 250 
Gaussian maps to model the CMB sky.   For the Gaussian maps we run synfast routine of Healpix with an
in-file representing the WMAP 5-year best-fit CMB anisotropy power spectrum and generate maps with
information out to $l = 600$.  We then use {\it anafast}, without employing an iteration scheme, masking with the $KQ75$ mask, to produce
$a_{lm}$'s for the Gaussian maps out to $l=600$. We will refer to these $a_{lm}$'s now collectively as $a_{lm}^{G}$.

\subsubsection{Noise}
\label{noise}

In addition to these Gaussian maps we create 250 noise maps for each
each of our frequency bands: Q, V and W.   We generate these maps from
white noise with mean = 0 and standard deviation = 1 taking into account
$\sigma_0$ and $N_{Obs}$ as follows:    
\begin{equation}
N(\hat{\mathbf n}) = {\sigma_0 \over \sqrt{N_{\rm obs}}} n(\hat{\mathbf n})
\end{equation}
where $N(\hat{\mathbf n})$ is our noise map and $n(\hat{\mathbf n})$ is
a map made of pure white noise, $N_{{\rm obs}}$ is the number of
observations per pixel and $\sigma_0$ is the rms noise per observation.
We use the frequency dependent $N_{{\rm obs}}$ for each point in the sky
provided by the WMAP 5-year Stokes-I map fits files and take $\sigma_0=2.197,\
3.133, \ $and 6.538 mK as established by the WMAP team for Q, V and W band 5-year data
respectively\cite{Jarosik:2006ib,Hinshaw:2008kr}. See also Table~\ref{tab:params}.

Starting with these noise maps we create alms using anafast with the
$KQ75$ mask with no iteration scheme out to $l=600$.  We will henceforth
refer to these alms collectively as $a_{lm}^N$. Furthermore, to
calculate the power spectrum from these noise maps we
use Healpix to evaluate the analytical expression:
\begin{equation}
\label{eq:Nl}
 N_l = \Omega_{\rm pix}\int \frac{d^2\hat{\mathbf n}}{4\pi f_{\rm
  sky}}\frac{\sigma_0^2M(\hat{\mathbf   n})}{N_{\rm obs}(\hat{\mathbf n})},
\end{equation}
where $\Omega_{\rm pix}\equiv 4\pi/N_{\rm pix}$ is the solid angle per
pixel, $M(\hat{\mathbf n})$ is the {\it KQ75} mask, $f_{\rm sky}=0.718$
is the fraction of sky retained by the {\it KQ75}
mask\cite{Komatsu:2008hk}.  Fig. \ref{fig:Nl} shows the power spectrum
from our simulated noise maps for each frequency compared to the analytical
values quoted by the WMAP 5-year team \cite{Komatsu:2008hk}.
For reference, the beam functions $b_l^i$ are plotted is Fig. \ref{fig:bl}.

\begin{figure}
    \begin{center}
      \includegraphics[scale=0.45]{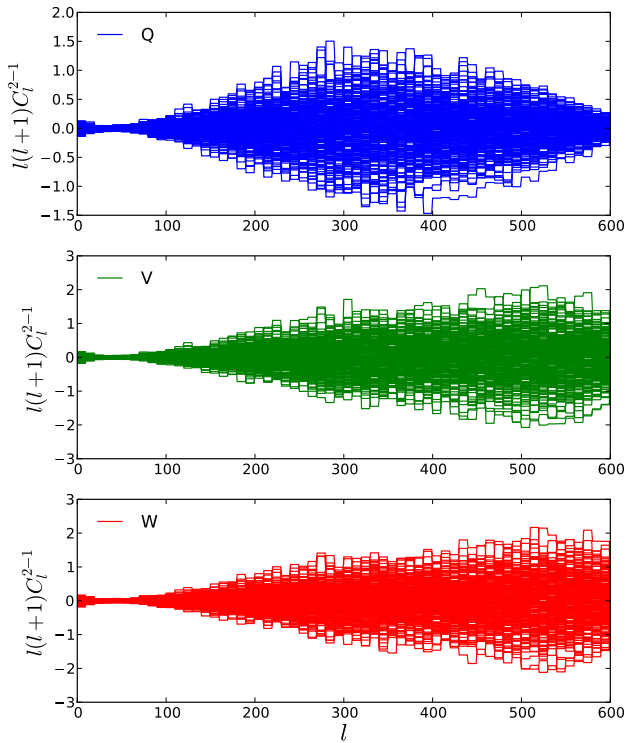} 
   \end{center}
   \vspace{-0.7cm}
   \caption[width=3in]{The results for $C_{l}^{2-1}$from all 250 simulations for the three frequency bands.  These plots have been binned with $\delta l = 10$}
   \label{fig:allsims}
\end{figure}

\begin{figure}
    \begin{center}
    \vspace{-0.6cm}
      \includegraphics[scale=0.45]{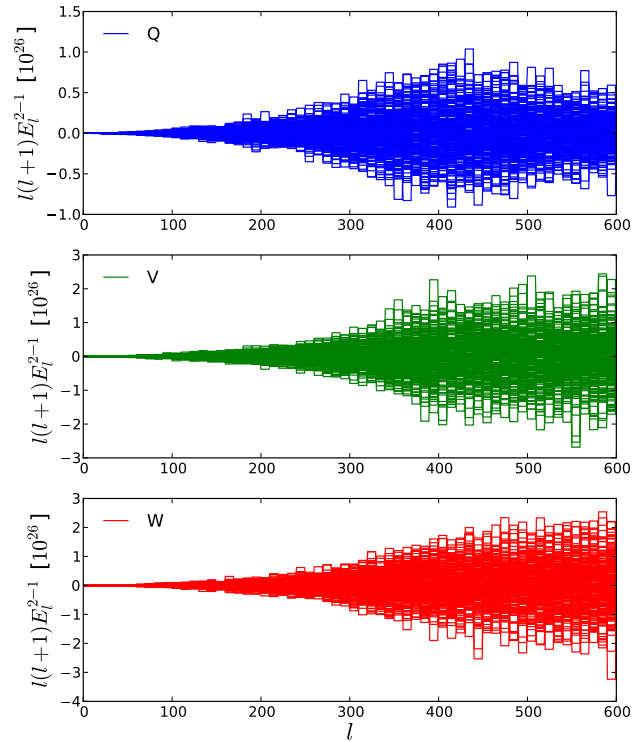} 
   \end{center}
   \vspace{-0.7cm}
   \caption[width=3in]{The results for $E_{l}^{2-1}$from all 250 simulations for the three frequency bands.  These plots have been binned with $\delta l = 10$}
   \label{fig:allsimsE}
\end{figure}

To use our estimator on the simulated maps we must add the noise to the
Gaussian maps while at the same time correcting for the beam.  To do this we
work in multipole space where we construct the total simulated $a_{lm}^S \equiv
a_{lm}^G b_l + a_{lm}^N$ where $b_l$ are the frequency dependent beam transfer
functions plotted in Fig. \ref{fig:bl}.

Figure \ref{fig:allsims} show the results of $C^{2-1}$ plotted with
respect to $l$ for each frequency band.  Similarly, Figure \ref{fig:allsimsE} show the all 250 simulated $E_{l}^{2-1}$ plotted for each frequency band.  These were binned with $\delta
l= 10$. 

From these 250 simulations we are able to develop a covariance matrix
that will be used for best fit estimates with error bars.  We find this covariance matrix by binning all 250 resulting estimators, $C_{l}^{2-1}$ (or $E_{l}^{2-1}$), in bins of $\delta l = 40$.  We can then treat each of these as an observation for each $l$ bin and create the covariance matrix by calculating the covariance of these observations.  This produced and $N$x$N$ covariance matrix where $N$ is the number of $l$ bins.

Figure \ref{fig:corr} shows the correlation matrices from the simulations.  These matrices were obtained by taking the covariance matrix, $C_{ij}$ and building the correlation matrix $\hat C_{ij}$ from the normalization:

\begin{figure}[h!]
    \begin{center}
      \includegraphics[scale=0.43]{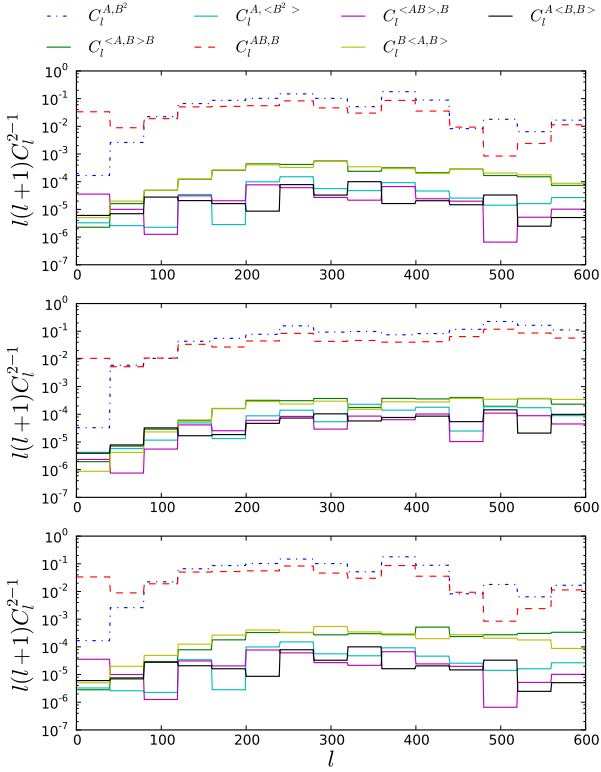} 
   \end{center}
   \vspace{-0.7cm}
   \caption[width=3in]{Contributions to $C_l^{1-2}$ for Q, V and W maps}
   \label{fig:tot}
\end{figure}

\begin{figure}[h!]
    \begin{center}
      \includegraphics[scale=0.43]{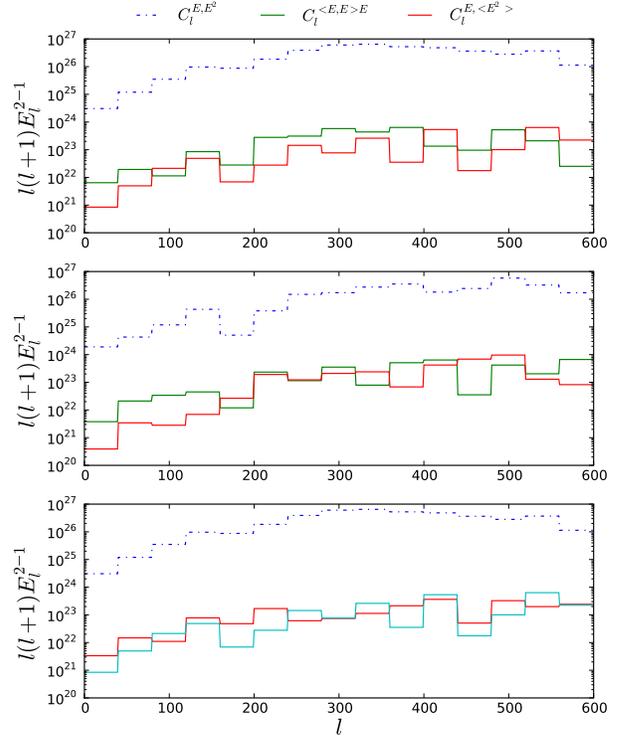} 
   \end{center}
   \vspace{-0.7cm}
   \caption[width=3in]{Contributions to $E_l^{1-2}$ for Q, V and W maps}
   \label{fig:totE}
\end{figure}

\begin{equation}
\hat C_{ij} = {C_{ij}\over \sqrt{C_{ii} C_{jj}}}
\end{equation}

We see that the correlation matrix obtained from the $C_{l}^{2-1}$ simulations show that these estimators have highly correlated $l$ bins.  It is interesting to note that the low $l$ bins are highly correlated with each other and the high $l$ bins are highly correlated with each other but low $l$ bins are not correlated strongly with the high ell bins.  

We also see correlation in the $E_{l}^{2-1}$ estimators but not nearly to as great a degree as $C_{l}^{2-1}$ above.  Furthermore, in the bottom of Figure \ref{fig:corr} we see the full $C_{l}^{2-1} + E_{l}^{2-1}$ correlation matrix and note there is correlation between the $l$ bins between $C_{l}^{2-1}$ and $E_{l}^{2-1}$ but not as much as there is between $C_{l}^{2-1}$ alone.

\subsection{Best Fit Estimation}

In order to fit the data, we use a least squares fitting analysis.  Given a
data set consisting of $n$ points $(x_i,y_i)$, we can fit this data with a
model function $f(\mathbf x,\mathbf p)$ where there are $m$ adjustable
parameters held in the vector ${\mathbf p}$.  We wish to find which of those
parameter values best fit the data.  

To do this we minimize the $\chi^{2}$ value defined as:
\begin{equation}
\chi^2  = (\mathbf y^T - \mathbf M \cdot \mathbf p)^T\mathbf C^{-1} (\mathbf y - \mathbf M \cdot \mathbf p).
\label{eqn:chi2}
\end{equation}
where $\mathbf y$ defines our data points we would like to fit to, $\mathbf p$ are the parameters we wish to solve for, $\mathbf M$ is a matrix containing our theoretical model we use for fitting and  $\mathbf C$ is our covariance matrix described above.  

For example, for a single frequency analysis where we would like to fit for $f_{\rm NL}$ and the coefficients for point sources: $\mathbf y = C_{l}^{2-1D}$ taken from data, $\mathbf M$ is the vector containing $\left<C_{l}^{2-1Th},PS^{Th}\right>$ and $\mathbf p = \left<f_{\rm NL},A_{i}\right>$ with $A_{i}$ being the coefficient for point sources.

We minimize $\chi^{2}$ by setting its derivative to zero and solving for $\mathbf p$ yielding:
\begin{equation}
{\mathbf  p} = ({\mathbf M^T \mathbf C^{-1} \mathbf M})^{-1}{\mathbf M^T \mathbf C^{-1} \cdot \mathbf y}.
\end{equation}

Lastly, we find the error bars for our best fit parameters via
\begin{equation}
\Delta \mathbf p^2 = (\mathbf M \mathbf C^{-1} \mathbf M)^{-1} 
\end{equation}
where the diagonal of this matrix gives the variance of the parameters and the $\chi^2$ fit is given by
equation~(\ref{eqn:chi2}).

\section{Results and Discussion}
\label{sec:results}

\subsection{$f_{\rm NL}$ estimate}

\begin{figure}
    \begin{center}
      \includegraphics[scale=0.45]{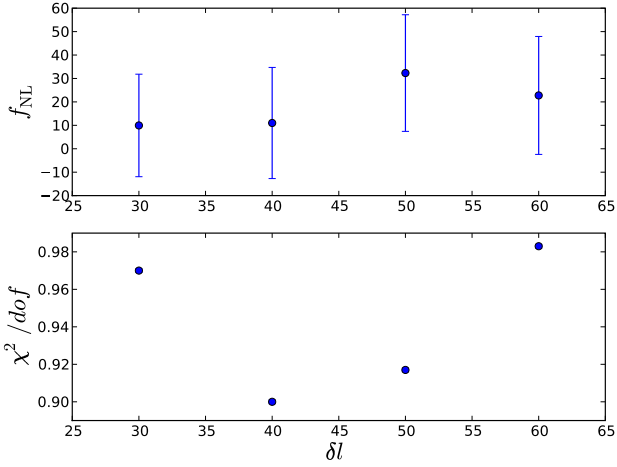} 
   \end{center}
   \caption[width=3in]{The various $f_{\rm NL}$ values taken from various binnings of the data from a full analysis considering both $C_l^{2-1}$ and $E_l^{2-1}$ with point sources and ISW.}
   \label{fig:fnlvbin}
\end{figure}

\begin{figure}
    \begin{center}
        \includegraphics[scale=0.45]{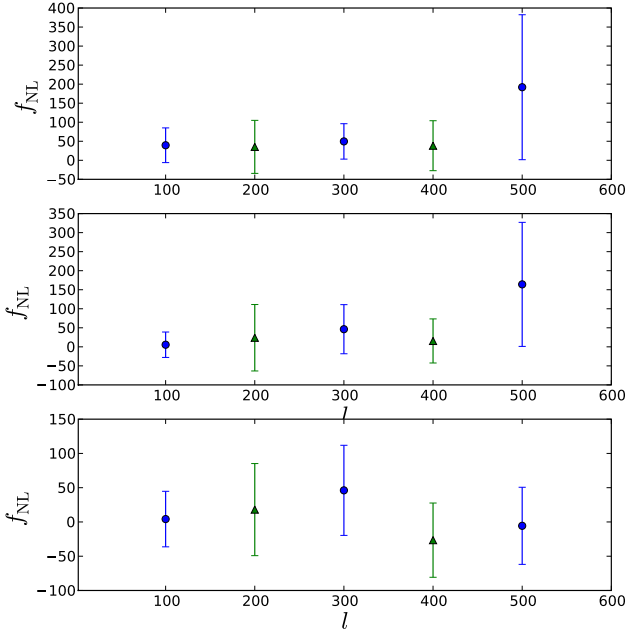} 
   \end{center}
   \caption[width=3in]{Angular dependance of $f_{\rm NL}$ between $2 < l < 600$ in bins of 200.  
Top is for $C_l^{2-1}$ only and with point sources.  Middle is the full measurement with $C_l^{2-1}$ and $E_l^{2-1}$ and using point sources.
The lower panel is full measurement with both point sources and lensing-secondary correlations.  The blue circles use mutually disjoint bins from each other.  The green triangles also use mutually disjoint bins.}
   \label{fig:angular}
\end{figure}

\begin{figure}
    \begin{center}
      \includegraphics[scale=0.45]{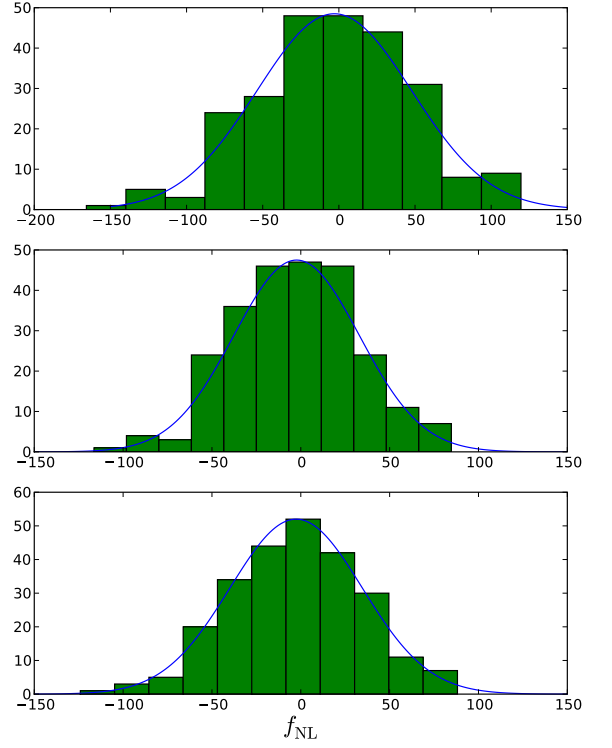} 
   \end{center}
   \vspace{-0.7cm}
   \caption[width=3in]{Histogram of the $f_{\rm NL}$ estimated from Gaussian and noise simulations for the cross-skewness statistic $S_3=\sum(2l+1)C_l^{2-1}$. Top: Q, Mid: V, Bottom: W.  A best fit Gaussian curve is plotted in blue over each histogram.}
   \label{fig:skewhist}
\end{figure}

We now discuss the results of our analysis. The primordial and foreground
non-Gaussianity parameter estimates are summarized in Table~\ref{tab:results1} for the case with and without point sources and
in Table~\ref{tab:results2} for the case with both point sources and lensing-secondary correlation.  
For each of these analyses, we bin and tabulate our measurements with bins of $\delta l = 40$. 
When $\delta l < 20$ the data are noisy to see the overall structure with a large covariance between adjacent bins and when
$\delta l > 100$ information from the fluctuating point source curves is lost leading to a large degeneracy
between parameters and an increase in parameter errors. 
Furthermore, of all binning widths between $20 < \delta l < 100$ the results are
similar, but the best $\chi^2$ value is always found with a binning at $\delta l = 40$ (Fig. \ref{fig:fnlvbin}).
Note that in the limit of a large $\delta l$ bin (with $\delta l > 200$) we
effectively reach the case of determining $f_{\rm NL}$ similar to the previous skewness statistic, with effectively just one data point per band.

In Table~II and III the first set of results, denoted by $C_l^{2-1}$, 
show the case when we fit our measured $C_l^{2-1}$ to the theoretical predictions involving a combination of primordial non-Gaussianity, point sources,
and lensing correlations as shown in Fig. \ref{fig:theoryC21}.   Given that Q map leads to a poor $\chi^2$ when model fitting Q alone or Q in combination
with other maps, we exclude the Q+V+W combination and instead consider V+W as our preferred set of maps.
When fitting to V and W, we compute the covariance of V and W, for example $\langle C_l^{2-1,V} C_l^{2-1,W} \rangle$ - $\langle C_l^{2-1,V} \rangle\langle C_l^{2-1,W} \rangle$.
Without point sources and lensing and simply fitting to $f_{\rm NL}$ with $C_l^{2-1}$ we find $4.8 \pm 27.7$\footnote{We quote 1$\sigma$ results with $\pm$ error and 2$\sigma$ result as a range.}. If the shot-noise from point sources are included, after marginalizing over $A_V$ and $A_W$, we find  $f_{\rm NL}=39.0 \pm 30.7$.

\begin{figure}[h!]
    \hspace{-3cm}
    \begin{center}
          \includegraphics[scale=0.4]{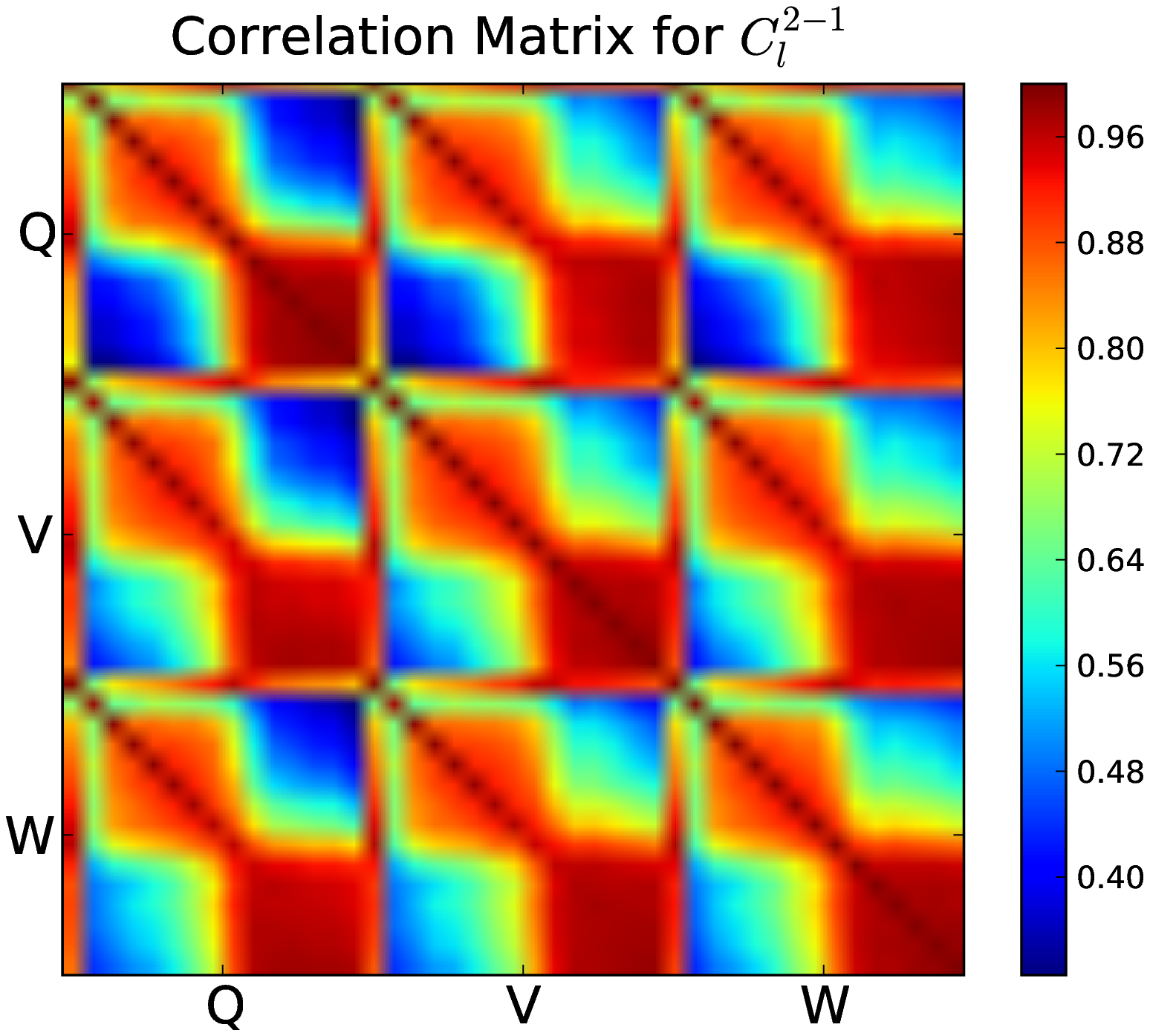} 
          \includegraphics[scale=0.4]{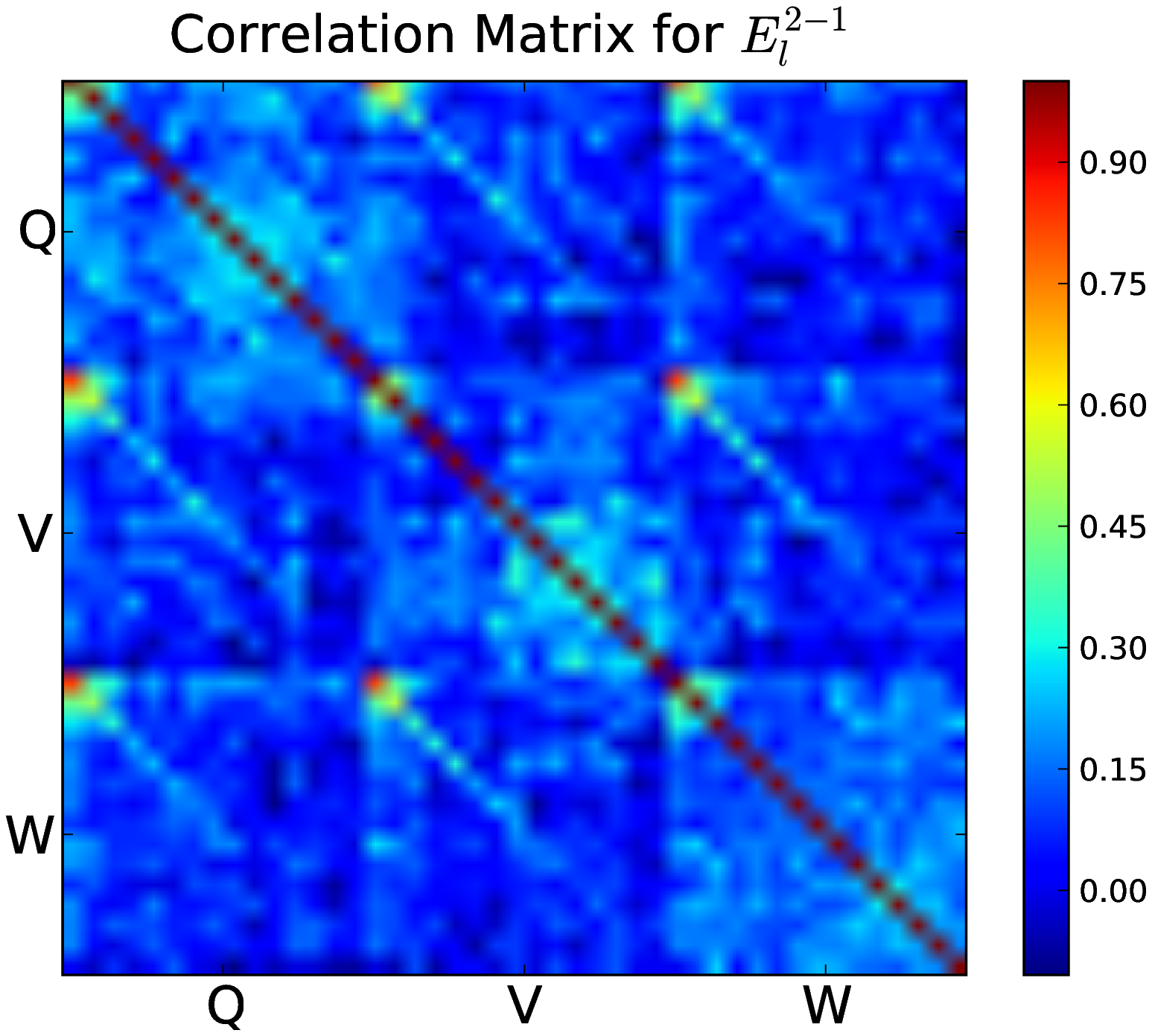} 
      \includegraphics[scale=0.4]{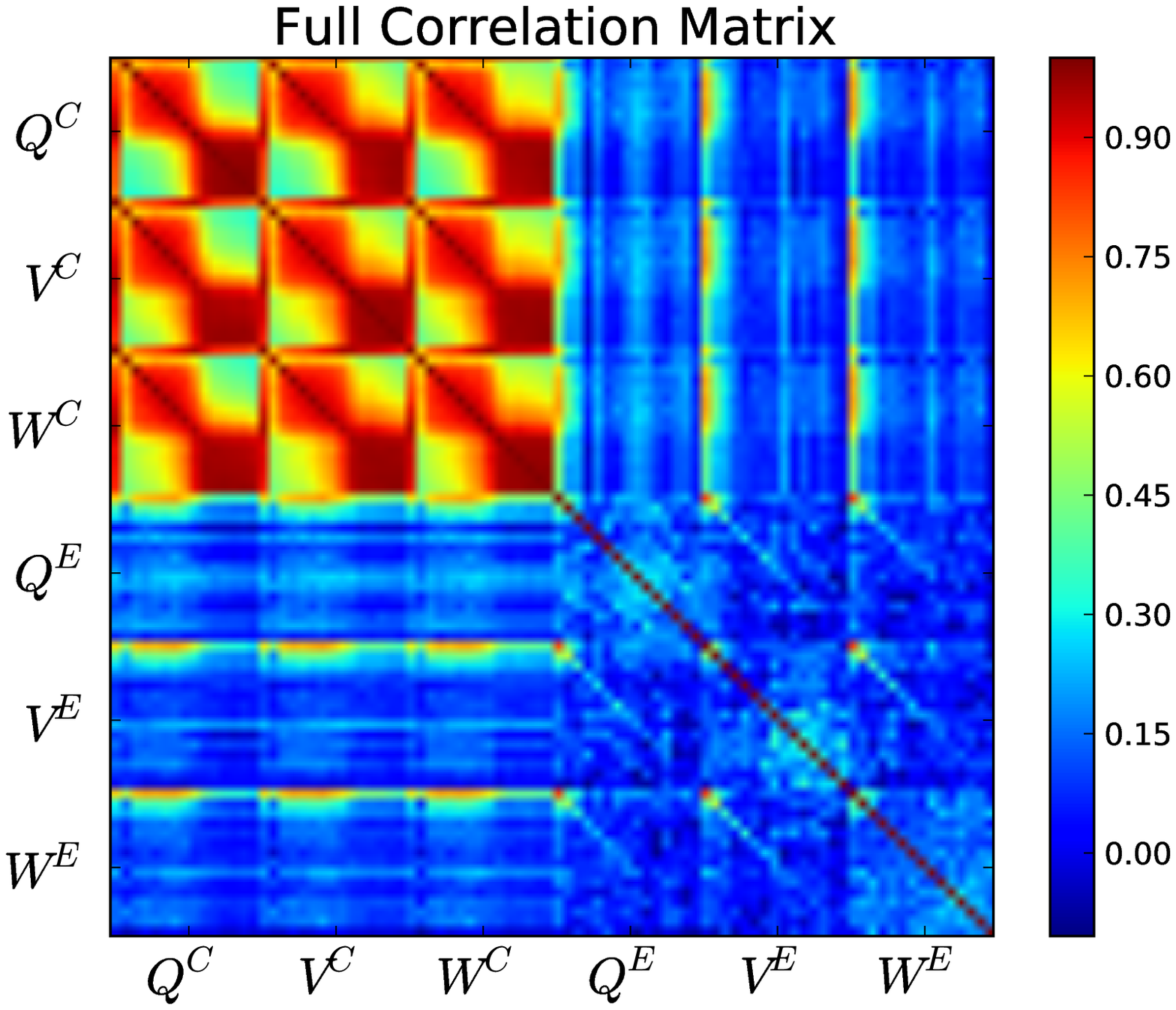} 
   \end{center}
   \vspace{-0.7cm}
   \caption[width=3in]{Matrices showing correlations between each frequency for both $C_{l}^{2-1}$ and $E_{l}^{2-1}$.  The upper left hand corner shows correlations between small ell, and moving toward the lower right corner shows correlations between high ell.}
   \label{fig:corr}
\end{figure}

As we discussed earlier, however,
fitting to $C_l^{2-1}$ alone with point sources lead to a worse determination of $f_{\rm NL}$ than the case where point sources are ignored due to the degeneracy between
primordial non-Gaussianity and point sources.  Thus, we also include E maps in our analysis with the associated
results from the skewness power spectrum denoted with $E^{2-1}$ in Table~\ref{tab:results1} and \ref{tab:results2}.  
The E maps provide a better estimator for the point sources but a worse
estimator for $f_{\rm NL}$ than the $C^{2-1}$ estimator, for reasons we discussed already. (Fig.
\ref{fig:theoryE21})  With $E_l^{2-1}$, the error bars for the point source amplitudes are about half of what
they were for $C^{2-1}$ alone, whereas the error bars on $f_{\rm NL}$ are about three times worse. 

One interesting thing to note is that $E_{l}^{2-1}$ is always positive.
This shows up in the best fit $f_{\rm NL}$ values, where for $C_{l}^{2-1}$ only the Q map pushes $f_{\rm NL}$ towards  a negative value, 
whereas $E_{l}^{2-1}$ from Q pushes $f_{\rm NL}$ to a large positive value.  In fact, if we include Q band and do a $f_{\rm NL}$ analysis with the E map alone, we find a 6$\sigma$ detection of the primordial
non-Gaussianity.  The $\chi^{2}$ from such an analysis, however, is poor and the result should not be trusted as a detection of a non-zero $f_{\rm NL}$.

Finally we consider the best fit when $C_l^{2-1}$ and $E_l^{2-1}$ are
combined.   The V+W analysis gives us the best constraint on $f_{\rm NL}$ with $-36.4 < f_{\rm NL} < 58.4$ at the 95\% confidence level or $(11.0 \pm 23.7)$ at the 68\% confidence level,
when we include both point sources and the lensing-secondary correlation and marginalize over $(A_V,A_W,\eta_V,\eta_W)$. 
As with the $C_{l}^{2-1}$ only analysis, this combined analysis has $f_{\rm NL}$ consistent with zero at 1$\sigma$.
As can be seen by comparing Table~II and III V+W case, our $f_{\rm NL}$ is essentially the same whether we include the lensing-secondary bispectrum or not.

While we do not include Q-band in our $f_{\rm NL}$ estimate, the
 Q-band point source amplitude of $(24.2 \pm 5.0) \times 10^{-25}$ sr$^2$ using the combination of $C_l^{2-1}$ and $E_l^{2-1}$ is
consistent with the WMAP team's preferred value for the point source amplitude of  $(4.3 \pm 1.3) \times 10^{-5}$ $\mu$K$^3$-sr$^2$ \cite{Komatsu:2008hk}. 
In their units, our $b_{\rm PS}^Q$ is equivalent to $(4.9 \pm 1.0) \times 10^{-5}$ $\mu$K$^3$-sr$^2$.
While we cannot make an exact comparison as WMAP team tabulates their point source values with $l_{\rm max}$ of 900, our values for $b_{\rm PS}^V$ and $b_{\rm PS}^W$ 
are also within uncertainties consistent with previous measurements. While the non-Gaussianity associated with point sources is detected,
we do not detect the lensing-secondary bispectrum. It is likely that the $C_l^{2-1}$ and $E_l^{2-1}$ are not the best ways to detect this correlation.
The best-fit values for $\eta_V$ and $\eta_W$, however, are close to their 1$\sigma$ errors.

As tabulated in Table~III, including the bispectrum of lensing-secondary correlations does not lead to a significant degradation of $f_{\rm NL}$ measurement. We find $f_{\rm NL}=11.0 \pm 23.7$ at the 1$\sigma$
confidence level, but we do not find a detection of $\eta_i$ in each of the three bands when $C_l^{2-1}$ is combined with $E_l^{2-1}$.

Note that our V+W  analysis gives   a $f_{\rm NL}$ value fully consistent with zero at the 1$\sigma$ level. Previous results have
suggested a marginal hint of a primordial non-Gaussianity with the most recent optimal anlaysis giving $f_{\rm NL}=38\pm21$ \cite{Smith:2009jr} (Table~V).
Compared to this result, our V+W has a slightly worse error with $11.0 \pm 23.7$, and the increase of 13\% is consistent with the fact that
our analysis is sub-optimal.   As discussed earlier, however, our approach is not different from
both the WMAP team's approach \cite{Komatsu:2008hk} and previous other
estimates of $f_{\rm NL}$~\cite{Yadav:2007yy}.  Moreover, our $l_{\rm max}$ is set at 600, while their analysis extends to 750.

To see if there is any scale dependence to non-Gaussianity we bin $f_{\rm NL}$ in widths of 200 and estimate the value between $0 < l < 600$.
The results are shown in Fig.~16 and tabulated in Table~IV. Except in the last bin for the case with point sources only between $400 < l < 600$, our $f_{\rm NL}$ values are
fully consistent with zero at the 1$\sigma$ level and the last bin is consistent with zero at the 2$\sigma$ level. The last bin also has a large error
due to the increase of the instrumental noise. For the same reason, we do not pursue a measurement of $f_{\rm NL}$ when $l > 600$.

It is also interesting to note how accurate our overall error estimate is. As we compute our covariances with 250 simulations there is an inherent 
 error of 1/$\sqrt{250}$ in the error bars we obtained in this analysis.  Because of this, we note that a more accurate estimate of $f_{\rm NL}$ 
should be to consider it as $11.0 \pm 23.7 (\pm 1.5)$ where the extra error within the bracket denotes an additional statistical error associated with
the finite number of simulations.

\begin{table*}[htbp]
  \centering
  \begin{tabular}{@{} |c|c|c|c|c|c|c| @{}}
    \hline
    Type & $f_{\rm NL}$ (no PSs) & $f_{\rm NL}$ (w/PSs) & $A_Q$ & $A_V$ & $A_W$ & $\chi^2/dof$ \\
    \hline
    $C_l^{2-1}$ &  &  &    &  &  & \\    
    Q & $-61.6 \pm 32.2$ & $-10.5 \pm 33.6$ &$62.0 \pm 12.1$ & &   & 1.6 \\
     V & $5.4 \pm 30.4$ & $36.5 \pm 32.9$  &  & $22.1 \pm 9.0$ &  &  0.6 \\
    W & $5.5 \pm 30.0$ & $31.8 \pm 33.3$  &  &  & $ 18.6 \pm 10.1$ &  0.6 \\
    V+W & $4.8 \pm 27.7$&$39.0 \pm 30.7$ &  & $18.5 \pm 8.2$ & $25.3 \pm 9.2$ & 1.0 \\
    \hline
    $E_l^{2-1}$ &  &  &   &  & &   \\
    Q & $426.4 \pm 100.5$ & $191.5 \pm 115.4$  & $57.0 \pm 13.8$ & &  & 1.3 \\
    V & $159.1 \pm 98.0$ & $94.2 \pm 106.6$  &  & $13.8 \pm 8.9$ &  & 0.3 \\
    W & $90.4 \pm  102.6$ & $49.2 \pm 112.4$  & & & $9.6 \pm 10.7$& 0.3 \\
    V+W  & $133.1 \pm 140.9$ & $69.8 \pm 100.6$  &  & $16.2 \pm 8.6$ & $9.4 \pm 10.3$ &  0.8 \\
    \hline
    Full &  &  &  &   &  & \\
    Q & $-23.1 \pm 29.4$ & $-22.0 \pm 29.4$ & $24.2 \pm 5.0$ & &  & 3.2 \\
    V & $13.1 \pm 26.8$ & $16.3 \pm 26.8$  &  & $4.2 \pm 2.1$ &  &   0.6 \\
    W & $19.5 \pm 26.9$ & $19.5 \pm 27.0$  &  & & $0.4 \pm 2.3$   & 0.6 \\
    V+W & $11.4 \pm 23.6$ &  $11.5 \pm 23.6$  &  & $5.0 \pm 1.8$& $-1.8 \pm 2.0$ & 0.9 \\
    \hline
  \end{tabular}
  \caption{
Parameter estimates with $C_l^{2-1}$ (top), $E_l^{2-1}$ (middle), and the combination of the two (bottom) with Q, V, W and V+W maps for the case where we
ignore point sources and including point sources. The point source amplitudes are listed under columns for $A_i$'s. 
The Q-band point source amplitude of
$(24.2 \pm 5.0) \times 10^{-25}$ sr$^2$, equivalent to $(4.9 \pm 1.0) \times 10^{-5}$ $\mu$K$^3$-sr$^2$ is consistent with the WMAP team's preferred value of 
$(4.3 \pm 1.3) \times 10^{-5}$ $\mu$K$^3$-sr$^2$. The value of $f_{\rm NL}$ with the amplitude of point sources marginalized over
$11.5 \pm 23.6$.}
  \label{tab:results1}
\end{table*}

\begin{table*}[htbp]
  \centering
  \begin{tabular}{@{} |c|c|c|c|c|c|c|c|c| @{}}
    \hline
    Type & $f_{\rm NL}$ (PS + lensing) & $A_Q$ & $A_V$ & $A_W$ & $\eta_Q$ & $\eta_V$ & $\eta_W$ & $\chi^2/dof$ \\
    \hline
    $C_{l}^{2-1}$   &  &  &  &  &  &  & & \\  
    Q & $21.1 \pm 40.3$ & $-80.2 \pm 39.3$ & &  & $-11.7 \pm 5.8$& &  & 3.4 \\
    V & $15.7 \pm 38.9$ &  & $8.7 \pm 23.0$ &  &  & $-3.7 \pm 4.6$ &  & 1.0 \\
    W & $-13.5 \pm 39.8$ &  &  & $ 39.7 \pm 25.6$ &  &  & $0.6 \pm 4.4$ & 1.2 \\
    V+W &$14.3 \pm 37.6$ &  & $18.2 \pm 20.8$ & $9.0 \pm 22.0$ &  & $-2.7 \pm 4.1$ & $-2.2 \pm 4.0$ & 1.3 \\
    \hline
    $E_{l}^{2-1}$  &  &  &  &  &  &  & & \\
        Q & $122.2 \pm 118.6$ & $8.5 \pm 6.2$ & &  & $6.6 \pm 1.7$& &  & 0.7 \\
    V & $80.5 \pm 107.8$ &  & $2.1 \pm 2.6$ &  &  & $1.2 \pm 1.1$ &  & 0.3 \\
    W & $62.3 \pm 113.2$ & &  & $-0.2 \pm 2.5$ & &  & $0.9 \pm 1.3$ & 0.3 \\
    V+W  & $72.0 \pm 103.1$ &  & $1.9 \pm 2.4$ & $-0.5 \pm 2.4$ &  & $1.4 \pm 1.1$ & $1.3 \pm 1.2$ & 0.8 \\
    \hline
    Full &  &  &  &  &  &  & & \\
    Q & $21.8 \pm 29.6$ & $24.0 \pm 5.7$ & &  & $0.2 \pm 1.2$& &  & 3.3 \\
    V & $16.7 \pm 27.1$ &  & $4.1 \pm 2.4$ &  &  & $0.2 \pm 0.5$ &  & 0.6 \\
    W & $18.7 \pm 27.2$  &  & & $0.5 \pm 2.3$   & & & $-0.3 \pm 1.0$ & 0.8 \\
    V+W &  $11.0 \pm 23.7$ & & $2.8 \pm 2.2$ &  $-0.4 \pm 2.2$ & & $1.0 \pm 0.8$  & $-0.6 \pm 0.9$ & 0.9 \\
    \hline
  \end{tabular}
  \caption{
Parameter estimates with $C_l^{2-1}$ (top), $E_l^{2-1}$ (middle), and the combination of the two (bottom) with Q, V, W and V+W maps for the case where we
account for both point sources and the amplitude of lensing-secondary bispectrum. The point source amplitudes are listed under columns for $A_i$'s,
while the amplitude of lensing-secondary signal is tabulated under $\eta_i$'s.  Our preferred value of $f_{\rm NL}$ with the amplitude of point sources and
the lensing-secondary signal marginalized over using V and W maps in combination is $11.0 \pm 23.7$.}
  \label{tab:results2}
\end{table*}

\begin{table*}[htbp]
  \centering
  \begin{tabular}{@{} |c|c|c| @{}}
    \hline
    Type & $f_{\rm NL}$ (with PSs) & $f_{\rm NL}$ (PSs + lensing-secondary) \\ 
    \hline
    $C_{l}^{2-1}$ &  &  \\ 
    $2 < l < 200$ & $39.5 \pm 45.6$ & $5.5 \pm 33.4$  \\
    
$100 < l < 300$ & $35.3 \pm 69.6$   & $23.9 \pm 87.3$   \\
    
$200 < l < 400$ & $49.6 \pm 46.5$ & $46.3 \pm 64.5$   \\ 
$300 < l < 500$ & $38.3 \pm 65.6$ & $15.5 \pm 57.8$   \\  
    
$400 < l < 600$ & $192.0 \pm 190.4$  & $164.1 \pm 162.9$   \\  
    \hline
    Full &  & \\ 
    
$2 < l < 200$ & $-9.2246 \pm 44.6$   & $4.2 \pm 40.5$  \\ 
    
$100 < l < 300$ & $-6.1 \pm 101.4$   & $18.0 \pm 67.2$ \\
    
$200 < l < 400$  & $64.5 \pm 74.0$  & $46.1 \pm 65.8$ \\
    
$300 < l < 500$ & $68.3 \pm 92.8$   & $-26.5 \pm 54.2$  \\
    
$400 < l < 600$ & $103.6 \pm 178.3$ & $-5.6 \pm 56.3$  \\ 
  
 \hline
  
\end{tabular}
  
\caption{Independent estimates of $f_{\rm NL}$ in bins of $\delta l=200$ between $2 < l < 600$.}  
\label{tab:results3}
\end{table*}

\begin{table*}[htbp]
  \centering
  \begin{tabular}{@{} |c|c|c| @{}}
    \hline
    Technique & $f_{\rm NL}$ & Ref\\
    \hline 
    WMAP 3-Year, Skewness & $87 \pm 30$ & \cite{Yadav:2007yy} \\
    WMAP 5-Year, Skewness & $51 \pm 30$ & \cite{Komatsu:2008hk} \\
    WMAP 5-Year, Minkowski Functions & $-57 \pm 61$ & \cite{Komatsu:2008hk} \\
    WMAP 5-year, Wavelets & $31 \pm 24.5$ & \cite{Curto:2009pv} \\
    WMAP 5-year, Needlets & $84 \pm 40$ & \cite{Rudjord:2009mh} \\
    WMAP 5-year, N-point PDF & $30 \pm 62$ & \cite{Vielva:2008wn} \\
    WMAP ISW-correlation & $236\pm127$ & \cite{Afshordi:2008ru} \\
    Large-scale structure bias & $20.5\pm24.8$ & \cite{Slosar:2008hx}\\
    WMAP 5-Year, Optimal Estimator  & $38 \pm 21$ & \cite{Smith:2009jr} \\
    WMAP 5-year, Skew-power spectrum & $11.0 \pm 23.7 (\pm 1.5)$ & this paper \\
    \hline 
  \end{tabular}
  \caption{Summary of recent results on $f_{\rm NL}$ measurements. Compared to the expectation from Cramer-Rao bound using the Fisher matrix estimate, our measurement is sub-optimal, but compared to the previous best estimate for $f_{\rm NL}$ of $38 \pm 21$, our estimate is fully consistent with zero at the 1$\sigma$ confidence level.}
  \label{tab:fnls}
\end{table*}

\subsection{Cross-Skewness}

Previous results for $f_{\rm NL}$ from the WMAP 5-year team compute $f_{\rm NL}$ by compressing all information into a single quantity called cross-skewness 
defined by equation~\ref{eq:skewness}.  To compare our measurement $C_{l}^{2-1}$ with their results we calculate our own equivalent version of this cross skewness statistic defined as
\begin{equation}
\hat{S}_{AB^2}=\sum (2l+1) C_l^{2-1,D}
\end{equation}
where $C_{l}^{2-1,D}$ is the estimator obtained from data. We also compute the skewness of the E map using $E_l^{2-1,D}$ in above. We jointly fit $\hat{S}_{AB^2}$ and $\hat{S}_{E^3}$
with a combination of $f_{\rm NL}$ and $A_i$ by effectively comparing the statistic from data to prediction from theory with theory expectation computed as, for example,
$S_{AB^2} = \sum (2l+1)C_l^{2-1,Th}$.
In order to determine the errors we also preform the same cross-skewness analysis on all 250 simulations and calculate the covariance of $\hat{S}_{AB^2}$ and $\hat{S}_{E^3}$ 
from these 250 numbers for each frequency.   We find that
$f_{\rm NL}$ estimated from each of the 250 Gaussian and noise simulations lead to
a Gaussian error distribution (Figure~\ref{fig:skewhist}).

We tabulate our results for $f_{\rm NL}$ after marginalizing over $A_i$'s in Table~VI. Here, when doing the summations 
we set $l_{\rm max}=500$ so we can compare directly with WMAP 5-year published results \cite{Komatsu:2008hk}. We see that for all three channels
we have good agreement with the WMAP team's 5-year findings. Our best-fit value tends to be bit more positive than quoted by the WMAP team (with $0.26\sigma,0.25\sigma,0.17\sigma$
in Q, V, and W respectively), but this is a small difference when compared to the
large error bar. The errors quoted in the WMAP 5-year paper is consistent with our measurements had we used the skewness statistic. However, as we discussed earlier, fitting to $C_l^{2-1}$
and $E_l^{2-1}$ leads to an improvement in the error estimate of $f_{\rm NL}$ since the shapes of the two skew spectra allow us to break the degeneracies better.
Comparing our V+W result using the two spectra to skewness for the same maps, we find that the improvement in the error is roughly  20\%.

\begin{table}[!t]
  \centering
  \begin{tabular}{@{} |c|c|c| @{}}
    \hline
     Band & $f_{\rm NL}$ & WMAP 5-year \\ 
    \hline
    Q & $-27.3 \pm 50.8$ & $-42 \pm 48$ \\ 
    V & $52.0 \pm 35.2$  & $41 \pm 35$\\ 
    W & $50.5 \pm 37.3$ & $46 \pm 35$\\ 
    \hline
  \end{tabular}
  \caption{Summary of results using the skewness where $S=\sum(2l+1)C_l^{2-1}$. Here 
we tabulate the values found in our analysis and the ones reported by the WMAP team \cite{Komatsu:2008hk}. We set $l_{\rm max}=500$ here.}
  \label{tab:label}
\end{table}

\section{Conclusion}
\label{sec:conclusion}

In this paper, we constrained the primordial non-Gaussianity parameter of the local model
$f_{\rm NL}$ using the skewness power spectrum associated with the two-to-one
cumulant correlator of cosmic microwave background temperature anisotropies.
This bispectrum-related skewness power spectrum was constructed after weighting the
temperature maps with the appropriate window functions to form an estimator
that probes the multipolar dependence of the underlying bispectrum associated
with primordial non-Gaussianity. 

We also estimate a separate skewness power spectrum more sensitive to unresolved point sources.   When compared
to previous attempts at measuring the primordial non-Gaussianity with WMAP
data, our estimators have the main advantage that we do not collapse
information to a single number. When model fitting  two-to-one skewness power spectrum
we make use of bispectra generated by primordial non-Gaussianity, radio point sources,
and lensing-secondary correlations. W

We analyze Q, V and W-band WMAP 5-year data using the KQ75 mask out to $l_{\rm
max}=600$.  Using V and W-band data and marginalizing over model parameters
related to point sources, our overall and preferred constraint on $f_{\rm NL}$
is 11.0 $\pm 23.7$ at the 68\% confidence level ($-36.4 < f_{\rm NL} < 58.4$ at
95$\%$ confidence). Despite previous claims,  we find no evidence for a non-zero
value of $f_{\rm NL}$ even marginally at the 1$\sigma$ level.


\acknowledgments
We are grateful to Eiichiro Komatsu, Dipak Munshi, and Kendrick Smith for assistance during
various stages of this work. J.S. acknowledges support from a GAANN fellowship.
Partial funding of AA and PS was from NSF CAREER AST-0645427.

\appendix

%
%


\begin{thebibliography}{99}

\bibitem{Guth:1980zm}
  A.~H.~Guth,
  Phys.\ Rev.\  D {\bf 23}, 347 (1981).

\bibitem{Linde:1981mu}
  A.~D.~Linde,
  Phys.\ Lett.\  B {\bf 108}, 389 (1982).

\bibitem{Albrecht:1982wi}
  A.~J.~Albrecht and P.~J.~Steinhardt,
  Phys.\ Rev.\ Lett.\  {\bf 48}, 1220 (1982).

\bibitem{Sato:1980yn}
  K.~Sato,
  Mon.\ Not.\ Roy.\ Astron.\ Soc.\  {\bf 195}, 467 (1981).

  
\bibitem{Kazanas:1980tx}
  D.~Kazanas,
  Astrophys.\ J.\  {\bf 241}, L59 (1980).


\bibitem{Starobinsky2}
  A.~A.~Starobinsky,
  JTEP Lett.\ {\bf 30}, 682 (1979).



\bibitem{GuthPi}
  A.~H.~Guth and S.~Y.~Pi,
  Phys.\ Rev.\ Lett.\  {\bf 49}, 1110 (1982).

\bibitem{Bardeen}
J.~M.~Bardeen, P.~J.~Steinhardt and M.~S.~Turner,
  Phys.\ Rev.\  D {\bf 28}, 679 (1983).

\bibitem{Hawking}
  S.~W.~Hawking,
  Phys.\ Lett.\  B {\bf 115}, 295 (1982).


\bibitem{Mukhanov}
  V.~F.~Mukhanov, H.~A.~Feldman and R.~H.~Brandenberger,
  Phys.\ Rept.\  {\bf 215}, 203 (1992).


\bibitem{Starobinsky}
  A.~A.~Starobinsky,
  Phys.\ Lett.\  B {\bf 117}, 175 (1982).

\bibitem{Komatsu:2008hk}
  E.~Komatsu {\it et al.}  [WMAP Collaboration],
  Astrophys.\ J.\ Suppl.\  {\bf 180}, 330 (2009)
  [arXiv:0803.0547 [astro-ph]].
  

\bibitem{Baumann:2008aq}
  D.~Baumann {\it et al.}  [CMBPol Study Team Collaboration],
  arXiv:0811.3919 [astro-ph].

\bibitem{Abbott}
  L.~F.~Abbott and M.~B.~Wise,
  Astrophys.\ J.\  {\bf 282}, L47 (1984).

\bibitem{Grishchuk:1974ny}
  L.~P.~Grishchuk,
  Sov.\ Phys.\ JETP {\bf 40}, 409 (1975)
  [Zh.\ Eksp.\ Teor.\ Fiz.\  {\bf 67}, 825 (1974)].



\bibitem{Planck}
    [Planck Collaboration],
  arXiv:astro-ph/0604069.

\bibitem{Bock:2006yf}
  J.~Bock {\it et al.},
  arXiv:astro-ph/0604101.


\bibitem{Bock:2009xw}
  J.~Bock {\it et al.}  [EPIC Collaboration],
  arXiv:0906.1188 [astro-ph.CO].

\bibitem{Bock:2008ww}
  J.~Bock {\it et al.},
  arXiv:0805.4207 [astro-ph].

\bibitem{Baumann:2008aj}
  D.~Baumann {\it et al.}  [CMBPol Study Team Collaboration],
  arXiv:0811.3911 [astro-ph].

\bibitem{Maldacena:2002vr}
  J.~M.~Maldacena,
  JHEP {\bf 0305}, 013 (2003)
  [arXiv:astro-ph/0210603].

\bibitem{Sasaki:1995aw}
  M.~Sasaki and E.~D.~Stewart,
  Prog.\ Theor.\ Phys.\  {\bf 95}, 71 (1996)
  [arXiv:astro-ph/9507001].

\bibitem{Lyth:2004gb}
  D.~H.~Lyth, K.~A.~Malik and M.~Sasaki,
  JCAP {\bf 0505}, 004 (2005)
  [arXiv:astro-ph/0411220].

\bibitem{Lyth:2005fi}
  D.~H.~Lyth and Y.~Rodriguez,
  Phys.\ Rev.\ Lett.\  {\bf 95}, 121302 (2005)
  [arXiv:astro-ph/0504045].


\bibitem{Acquaviva:2002ud}
  V.~Acquaviva, N.~Bartolo, S.~Matarrese and A.~Riotto,
  Nucl.\ Phys.\  B {\bf 667}, 119 (2003)
  [arXiv:astro-ph/0209156].

\bibitem{Allen87}
  T.~J.~Allen, B.~Grinstein and M.~B.~Wise,
  Phys.\ Lett.\  B {\bf 197}, 66 (1987).

\bibitem{Falk93}
  T.~Falk, R.~Rangarajan and M.~Srednicki,
  Astrophys.\ J.\  {\bf 403}, L1 (1993)
  [arXiv:astro-ph/9208001].

\bibitem{Salopek:1990jq}
  D.~S.~Salopek and J.~R.~Bond,
  Phys.\ Rev.\  D {\bf 42}, 3936 (1990).

\bibitem{Gangui94}
  A.~Gangui, F.~Lucchin, S.~Matarrese and S.~Mollerach,
  Astrophys.\ J.\  {\bf 430}, 447 (1994)
  [arXiv:astro-ph/9312033].


\bibitem{Mollerach:1989hu}
  S.~Mollerach,
  Phys.\ Rev.\  D {\bf 42}, 313 (1990).

\bibitem{Dvali:2003em}
  G.~Dvali, A.~Gruzinov and M.~Zaldarriaga,
  Phys.\ Rev.\  D {\bf 69}, 023505 (2004)
  [arXiv:astro-ph/0303591].

\bibitem{Linde:1984ti}
  A.~D.~Linde,
  JETP Lett.\  {\bf 40}, 1333 (1984)
  [Pisma Zh.\ Eksp.\ Teor.\ Fiz.\  {\bf 40}, 496 (1984)].

\bibitem{Berera:1995ie}
  A.~Berera,
  Phys.\ Rev.\ Lett.\  {\bf 75}, 3218 (1995)
  [arXiv:astro-ph/9509049].

\bibitem{ArkaniHamed:2003uy}
  N.~Arkani-Hamed, H.~C.~Cheng, M.~A.~Luty and S.~Mukohyama,
  JHEP {\bf 0405}, 074 (2004)
  [arXiv:hep-th/0312099].

\bibitem{Silverstein:2003hf}
  E.~Silverstein and D.~Tong,
  Phys.\ Rev.\  D {\bf 70}, 103505 (2004)
  [arXiv:hep-th/0310221].

\bibitem{Chen:2005fe}
  X.~Chen,
  Phys.\ Rev.\  D {\bf 72}, 123518 (2005)
  [arXiv:astro-ph/0507053].

\bibitem{Bartolo:2004if}
  N.~Bartolo, E.~Komatsu, S.~Matarrese and A.~Riotto,
  Phys.\ Rept.\  {\bf 402}, 103 (2004)
  [arXiv:astro-ph/0406398].

  
\bibitem{Komatsu:2001rj}
  E.~Komatsu and D.~N.~Spergel,
  Phys.\ Rev.\  D {\bf 63}, 063002 (2001)
  [arXiv:astro-ph/0005036].

\bibitem{Yadav:2007yy}
  A.~P.~S.~Yadav and B.~D.~Wandelt,
  Phys.\ Rev.\ Lett.\  {\bf 100}, 181301 (2008)
  [arXiv:0712.1148 [astro-ph]].

\bibitem{Smith:2009jr}
  K.~M.~Smith, L.~Senatore and M.~Zaldarriaga,
  arXiv:0901.2572 [astro-ph].



\bibitem{Serra}
  P.~Serra and A.~Cooray,
  Phys.\ Rev.\  D {\bf 77}, 107305 (2008)
  [arXiv:0801.3276 [astro-ph]].

\bibitem{Cooray:2001ps}
  A.~Cooray,
  Phys.\ Rev.\  D {\bf 64}, 043516 (2001)
  [arXiv:astro-ph/0105415].

\bibitem{Munshi:2009ik}
  D.~Munshi and A.~Heavens,
  arXiv:0904.4478 [astro-ph.CO].

\bibitem{Goldberg}
  D.~M.~Goldberg and D.~N.~Spergel,
  Phys.\ Rev.\  D {\bf 59}, 103002 (1999)
  [arXiv:astro-ph/9811251].


\bibitem{CooHu}
  A.~Cooray and W.~Hu,
  Astrophys.\ J.\  {\bf 548}, 7 (2001)
  [arXiv:astro-ph/0004151].


\bibitem{Hinetal95}
        G. Hinshaw, A. J. Banday, C. L. Bennett, K. M. Gorski and
        A. Kogut, \ApJ, 446, 67 (1995)


\bibitem{Feretal98}
        P. G. Ferreira, J. Magueijo and K. M. Gorksi, \ApJ, 503, 1
(1998); for updates, see also,
        J. Pando, D. Vallas-Gabaud D.  and L. Fang, \PRL, 79, 1611 (1998);
        A. J. Banday, S. Zaroubi, S. and K. M. Gorski, \ApJ in press
(astro-ph/9908070); B. Bromley and M. Tegmark, \ApJL, 524, L79 (1999)

\bibitem{Cooetal00}
        A. Cooray, W. Hu and  M. Tegmark, Astrophys. J., 540, 1 (2000)


\bibitem{SelZal96}
  U. Seljak and M. Zaldarriaga,
        \ApJ, 469, 437 (1996)

\bibitem{Serra2}
  P.~Serra, A.~Cooray, A.~Amblard, L.~Pagano and A.~Melchiorri,
  Phys.\ Rev.\  D {\bf 78}, 043004 (2008)
  [arXiv:0806.1742 [astro-ph]].





\bibitem{CooShe}
  A.~Cooray and R.~K.~Sheth,
  Phys.\ Rept.\  {\bf 372}, 1 (2002)
  [arXiv:astro-ph/0206508].

\bibitem{Cooray1}
  A.~Cooray,
  Phys.\ Rev.\  D {\bf 62}, 103506 (2000)
  [arXiv:astro-ph/0005287].

\bibitem{Cooray2}
  A.~Cooray,
  Phys.\ Rev.\  D {\bf 64}, 063514 (2001)
  [arXiv:astro-ph/0105063].


\bibitem{Cooray3}
  A.~Cooray,
  Phys.\ Rev.\  D {\bf 65}, 103510 (2002)
  [arXiv:astro-ph/0112408].

\bibitem{Afshordi}
  N.~Afshordi,
  Phys.\ Rev.\  D {\bf 70}, 083536 (2004)
  [arXiv:astro-ph/0401166].



\bibitem{KS}
  E.~Komatsu and U.~Seljak,
  Mon.\ Not.\ Roy.\ Astron.\ Soc.\  {\bf 336}, 1256 (2002)
  [arXiv:astro-ph/0205468].


\bibitem{Verde}
  A.~Mangilli and L.~Verde,
  arXiv:0906.2317 [astro-ph.CO].

\bibitem{Hanson}
  D.~Hanson, K.~M.~Smith, A.~Challinor and M.~Liguori,
  arXiv:0905.4732 [astro-ph.CO].

\bibitem{Cooray4}
  A.~Cooray, D.~Sarkar and P.~Serra,
  Phys.\ Rev.\  D {\bf 77}, 123006 (2008)
  [arXiv:0803.4194 [astro-ph]].

\bibitem{Veneziani}
  M.~Veneziani {\it et al.},
  arXiv:0904.4313 [astro-ph.CO].



\bibitem{Szapudi}
  G.~Chen and I.~Szapudi,
  Astrophys.\ J.\  {\bf 647}, L87 (2006)
  [arXiv:astro-ph/0606394].

\bibitem{KSW}
  E.~Komatsu, D.~N.~Spergel and B.~D.~Wandelt,
  Astrophys.\ J.\  {\bf 634}, 14 (2005)
  [arXiv:astro-ph/0305189].

\bibitem{Gorski:2004by}
  K.~M.~Gorski, E.~Hivon, A.~J.~Banday, B.~D.~Wandelt, F.~K.~Hansen, M.~Reinecke and M.~Bartelman,
  Astrophys.\ J.\  {\bf 622}, 759 (2005)
  [arXiv:astro-ph/0409513].


\bibitem{Jarosik:2006ib}
  N.~Jarosik {\it et al.}  [WMAP Collaboration],
  Astrophys.\ J.\ Suppl.\  {\bf 170}, 263 (2007)
  [arXiv:astro-ph/0603452].

\bibitem{Hinshaw:2008kr}
  G.~Hinshaw {\it et al.}  [WMAP Collaboration],
  Astrophys.\ J.\ Suppl.\  {\bf 180}, 225 (2009)
  [arXiv:0803.0732 [astro-ph]].

\bibitem{Curto:2009pv}
  A.~Curto, E.~Martinez-Gonzalez and R.~B.~Barreiro,
  arXiv:0902.1523 [astro-ph.CO].

\bibitem{Rudjord:2009mh}
  O.~Rudjord, F.~K.~Hansen, X.~Lan, M.~Liguori, D.~Marinucci and S.~Matarrese,
  arXiv:0901.3154 [astro-ph.CO].

\bibitem{Vielva:2008wn}
  P.~Vielva and J.~L.~Sanz,
  arXiv:0812.1756 [astro-ph].

\bibitem{Afshordi:2008ru}
  N.~Afshordi and A.~J.~Tolley,
  Phys.\ Rev.\  D {\bf 78}, 123507 (2008)
  [arXiv:0806.1046 [astro-ph]].

\bibitem{Slosar:2008hx}
  A.~Slosar, C.~Hirata, U.~Seljak, S.~Ho and N.~Padmanabhan,
  JCAP {\bf 0808}, 031 (2008)
  [arXiv:0805.3580 [astro-ph]].



\end{thebibliography}
\end{document}